\documentclass[journal,draftcls,onecolumn,12pt,romanappendices]{IEEEtran-v17}

\pdfoutput=1

\usepackage{xspace}
\usepackage{url}
\usepackage[cmex10]{amsmath}
\usepackage{bbm}
\usepackage{graphicx}
\usepackage{tikz}
\usepackage{pgfplots}
\usepackage{ifthen}

\newboolean{conf}
\setboolean{conf}{false} 

\newboolean{singcol}
\setboolean{singcol}{true}

\usepackage{fancyref} 
\usepackage{amsmath}
\usepackage{amssymb}
\usepackage{dsfont}
\usepackage{footnote}
\usepackage[nosort]{cite}
\usepackage[british]{babel}

\usepackage{color}
\definecolor{links}{rgb}{0.7,0,0}   
\definecolor{urls}{rgb}{0,0,0.8}    
\definecolor{cites}{rgb}{0,0,0.8}   

\usepackage{enumerate}

\usepackage[colorlinks,hyperindex,linkcolor=links,citecolor=cites,urlcolor=urls]{hyperref}

\usepackage{bm}
\usepackage{upgreek}
\usepackage{amssymb}
\usepackage{amsfonts}
\usepackage{mathrsfs}
\usepackage{xspace}

\makeatletter
\def\amsbb{\use@mathgroup \M@U \symAMSb}
\makeatother

\newcommand{\lefto}{\mathopen{}\left}
\newcommand{\safemath}[2]{\newcommand{#1}{\ensuremath{#2}\xspace}}
\safemath{\opE}{\amsbb{E}}
\newcommand{\Ex}[2]{\ensuremath{\amsbb{E}_{#1}\mathopen{}\left[#2\right]}} 	
\safemath{\prob}{\amsbb{P}}
\safemath{\bigO}{\mathcal{O}}
\safemath{\littleo}{\mathit{o}}

\safemath{\extendreal}{\overline{\realset}}

\newtheorem{thm}{Theorem}
\newtheorem{lemma}[thm]{Lemma}

\newtheorem{rem}{Remark}
\newtheorem{cond}{Condition}
\newtheorem{prop}[thm]{Proposition}

\newtheorem{cor}[thm]{Corollary}

\safemath{\matA}{\mathsf{A}}
\safemath{\matB}{\mathsf{B}}
\safemath{\matC}{\mathsf{C}}
\safemath{\matD}{\mathsf{D}}
\safemath{\matE}{\mathsf{E}}
\safemath{\matF}{\mathsf{F}}
\safemath{\matG}{\mathsf{G}}
\safemath{\matH}{\mathsf{H}}
\safemath{\matI}{\mathsf{I}}
\safemath{\matJ}{\mathsf{J}}
\safemath{\matK}{\mathsf{K}}
\safemath{\matL}{\mathsf{L}}
\safemath{\matM}{\mathsf{M}}
\safemath{\matN}{\mathsf{N}}
\safemath{\matO}{\mathsf{O}}
\safemath{\matP}{\mathsf{P}}
\safemath{\matQ}{\mathsf{Q}}
\safemath{\matR}{\mathsf{R}}
\safemath{\matS}{\mathsf{S}}
\safemath{\matT}{\mathsf{T}}
\safemath{\matU}{\mathsf{U}}
\safemath{\matV}{\mathsf{V}}
\safemath{\matW}{\mathsf{W}}
\safemath{\matX}{\mathsf{X}}
\safemath{\matY}{\mathsf{Y}}
\safemath{\matZ}{\mathsf{Z}}
\safemath{\matSigma}{\mathsf{\Sigma}}
\safemath{\matPhi}{\mathsf{\Phi}}
\safemath{\matLambda}{\mathsf{\Lambda}}
\safemath{\matDelta}{\mathsf{\Delta}}

\safemath{\randveca}{\bm{A}}
\safemath{\randvecb}{\bm{B}}
\safemath{\randvecc}{\bm{C}}
\safemath{\randvecd}{\bm{D}}
\safemath{\randvece}{\bm{E}}
\safemath{\randvecf}{\bm{F}}
\safemath{\randvecg}{\bm{G}}
\safemath{\randvech}{\bm{H}}
\safemath{\randveci}{\bm{I}}
\safemath{\randvecj}{\bm{J}}
\safemath{\randveck}{\bm{K}}
\safemath{\randvecl}{\bm{L}}
\safemath{\randvecm}{\bm{M}}
\safemath{\randvecn}{\bm{N}}
\safemath{\randveco}{\bm{O}}
\safemath{\randvecp}{\bm{P}}
\safemath{\randvecq}{\bm{Q}}
\safemath{\randvecr}{\bm{R}}
\safemath{\randvecs}{\bm{S}}
\safemath{\randvect}{\bm{T}}
\safemath{\randvecu}{\bm{U}}
\safemath{\randvecv}{\bm{V}}
\safemath{\randvecw}{\bm{W}}
\safemath{\randvecx}{\bm{X}}
\safemath{\randvecy}{\bm{Y}}
\safemath{\randvecz}{\bm{Z}}

\safemath{\randvecLambda}{\bm{\Lambda}}

\safemath{\randmatA}{\amsbb{A}}
\safemath{\randmatB}{\amsbb{B}}
\safemath{\randmatC}{\amsbb{C}}
\safemath{\randmatD}{\amsbb{D}}
\safemath{\randmatE}{\amsbb{E}}
\safemath{\randmatF}{\amsbb{F}}
\safemath{\randmatG}{\amsbb{G}}
\safemath{\randmatH}{\amsbb{H}}
\safemath{\randmatI}{\amsbb{I}}
\safemath{\randmatJ}{\amsbb{J}}
\safemath{\randmatK}{\amsbb{K}}
\safemath{\randmatL}{\amsbb{L}}
\safemath{\randmatM}{\amsbb{M}}
\safemath{\randmatN}{\amsbb{N}}
\safemath{\randmatO}{\amsbb{O}}
\safemath{\randmatP}{\amsbb{P}}
\safemath{\randmatQ}{\amsbb{Q}}
\safemath{\randmatR}{\amsbb{R}}
\safemath{\randmatS}{\amsbb{S}}
\safemath{\randmatT}{\amsbb{T}}
\safemath{\randmatU}{\amsbb{U}}
\safemath{\randmatV}{\amsbb{V}}
\safemath{\randmatW}{\amsbb{W}}
\safemath{\randmatX}{\amsbb{X}}
\safemath{\randmatY}{\amsbb{Y}}
\safemath{\randmatZ}{\amsbb{Z}}
\safemath{\randmatSigma}{\mathbb{\Sigma}}
\safemath{\randmatPhi}{\mathbb{\Phi}}

\safemath{\pdff}{f}
\safemath{\pdfp}{p}
\safemath{\pdfq}{q}

\safemath{\cdfF}{F}
\safemath{\cdfP}{P}
\safemath{\cdfQ}{Q}

\safemath{\veca}{\bm{a}}
\safemath{\vecb}{\bm{b}}
\safemath{\vecc}{\bm{c}}
\safemath{\vecd}{\bm{d}}
\safemath{\vece}{\bm{e}}
\safemath{\vecf}{\bm{f}}
\safemath{\vecg}{\bm{g}}
\safemath{\vech}{\bm{h}}
\safemath{\veci}{\bm{i}}
\safemath{\vecj}{\bm{j}}
\safemath{\veck}{\bm{k}}
\safemath{\vecl}{\bm{l}}
\safemath{\vecm}{\bm{m}}
\safemath{\vecn}{\bm{n}}
\safemath{\veco}{\bm{o}}
\safemath{\vecp}{\bm{p}}
\safemath{\vecq}{\bm{q}}
\safemath{\vecr}{\bm{r}}
\safemath{\vecs}{\bm{s}}
\safemath{\vect}{\bm{t}}
\safemath{\vecu}{\bm{u}}
\safemath{\vecv}{\bm{v}}
\safemath{\vecw}{\bm{w}}
\safemath{\vecx}{\bm{x}}
\safemath{\vecy}{\bm{y}}
\safemath{\vecz}{\bm{z}}

\safemath{\veclambda}{\bm{\lambda}}
\safemath{\vecpi}{\bm{\pi}}
\safemath{\vecsigma}{\bm\sigma}              			

\safemath{\setA}{\mathcal{A}}
\safemath{\setB}{\mathcal{B}}
\safemath{\setC}{\mathcal{C}}
\safemath{\setD}{\mathcal{D}}
\safemath{\setE}{\mathcal{E}}
\safemath{\setF}{\mathcal{F}}
\safemath{\setG}{\mathcal{G}}
\safemath{\setH}{\mathcal{H}}
\safemath{\setI}{\mathcal{I}}
\safemath{\setJ}{\mathcal{J}}
\safemath{\setK}{\mathcal{K}}
\safemath{\setL}{\mathcal{L}}
\safemath{\setM}{\mathcal{M}}
\safemath{\setN}{\mathcal{N}}
\safemath{\setO}{\mathcal{O}}
\safemath{\setP}{\mathcal{P}}
\safemath{\setQ}{\mathcal{Q}}
\safemath{\setR}{\mathcal{R}}
\safemath{\setS}{\mathcal{S}}
\safemath{\setT}{\mathcal{T}}
\safemath{\setU}{\mathcal{U}}
\safemath{\setV}{\mathcal{V}}
\safemath{\setW}{\mathcal{W}}
\safemath{\setX}{\mathcal{X}}
\safemath{\setY}{\mathcal{Y}}
\safemath{\setZ}{\mathcal{Z}}
\safemath{\emptySet}{\varnothing}

\safemath{\veczero}{\mathbf{0}} 

\safemath{\diag}{\mathrm{diag}}

\safemath{\jpg}{\mathcal{CN}}			
\safemath{\complexset}{\amsbb{C}}
\safemath{\realset}{\amsbb{R}}
\safemath{\natunum}{\mathbb{N}}
\safemath{\posrealset}{\realset_{+}}
\safemath{\integerset}{\amsbb{N}}

\safemath{\define}{\triangleq}			

\usepackage{bm}
\usepackage{upgreek}
\usepackage{amssymb}
\usepackage{amsfonts}
\usepackage{mathrsfs}
\usepackage{xspace}

\safemath{\mi}{I}
\safemath{\difent}{h}

\safemath{\constrm}{\mathrm{const}}

\safemath{\NonnegReal}{\mathbb{R}^{+}}
\safemath{\re}{\mathrm{re}}
\safemath{\Real}{\mathrm{Re}} 
\safemath{\gradient}{\nabla}
\safemath{\genericpdf}{f}

\safemath{\bl}{n} 
\safemath{\error}{\epsilon} 

\safemath{\cohtime}{n_\mathrm{c}}
\safemath{\rxant}{m_\mathrm{r}}
\safemath{\txant}{m_\mathrm{t}}

\safemath{\snr}{P}
\safemath{\const}{k}

\safemath{\spanm}{\mathrm{span}}

\safemath{\altg}{\tilde{g}}

\safemath{\altk}{\tilde{k}}
\safemath{\altconst}{\altk}
\safemath{\altvecLambda}{\widetilde{\randvecLambda}}

\safemath{\altveclambda}{\tilde{\veclambda}}
\safemath{\altvecsigma}{\tilde{\vecsigma}}              

\safemath{\altLambda}{\widetilde{\Lambda}}

\safemath{\altgamma}{\tilde{\gamma}}
\safemath{\altsigma}{\tilde{\sigma}}      				
\safemath{\altdelta}{\tilde{\delta}}
\safemath{\altrho}{\tilde{\rho}}
\safemath{\altlambda}{\tilde{\lambda}}

\safemath{\altsnr}{\altrho}						

\safemath{\altmatSigma}{\matDelta}
\safemath{\altmatPhi}{\widetilde{\matPhi}} 		
\safemath{\inpdist}{Q_{\randmatX}}

\def\cohtime{n_{\mathrm{c}}}
\def\realg{\mathcal{N}}
\def\msg{M}
\def\deg{\mathrm{deg}}

\begin{document}
\IEEEoverridecommandlockouts

\title{\huge State-Dependent Gaussian Multiple Access Channels: New Outer Bounds  and Capacity Results}


\author{Wei Yang, \IEEEmembership{Member, IEEE}, Yingbin Liang, \IEEEmembership{Senior Member, IEEE}, \\Shlomo Shamai (Shitz) \IEEEmembership{Fellow, IEEE},  and H. Vincent Poor, \IEEEmembership{Fellow, IEEE}
\thanks{The work of W.~Yang and H. V. Poor was supported by the U. S. National Science Foundation under Grants ECCS-1343210 and ECCS-1647198. 
The work of Y. Liang was supported by the U. S. National Science Foundation under Grant CCF-1618127.
The work of S. Shamai  (Shitz) was supported by the European Union's Horizon 2020 Research And Innovation Programme, under grant agreement no. 694630.
The material of this paper will be presented in part at the IEEE International Symposium on Information Theory (ISIT), Aachen, Germany, June 2017.}
\thanks{W. Yang and H. V. Poor are with the Department of Electrical Engineering, Princeton University, Princeton, NJ 08544 USA (email: $\{$weiy, poor$\}$@princeton.edu).}
\thanks{Y. Liang is with the Department of Electrical Engineering and Computer Science, Syracuse University, Syracuse, NY 13244 USA (email: yliang06@syr.edu).}
\thanks{S. Shamai (Shitz) is with the Department of Electrical Engineering, Technion--Israel Institute of Technology, Technion City, Haifa 32000, Israel (email: sshlomo@ee.technion.ac.li).}
}


\maketitle
\begin{abstract}
This paper studies a two-user state-dependent Gaussian multiple-access channel (MAC) with state noncausally known at one encoder.  
Two scenarios are considered: i) each user wishes to communicate an independent message to the common receiver, and ii) the two encoders send a common message to the receiver and the non-cognitive encoder (i.e., the encoder that does not know the state) sends an independent individual message (this model is also known as the MAC with degraded message sets). 
For both scenarios, new outer bounds on the capacity region are derived, which improve uniformly over the best known outer bounds. 
In the first scenario, the two corner points of the capacity region as well as the sum rate capacity are established, and it is shown that a single-letter solution is adequate to achieve both the corner points and the sum rate capacity.
Furthermore, the full capacity region is characterized in situations in which the sum rate capacity is equal to the capacity of the helper problem.
The proof exploits the optimal-transportation idea of Polyanskiy and Wu (which was used previously to establish an outer bound on the capacity region of the interference channel) and the worst-case Gaussian noise result for the case in which the input and the noise are dependent.
\end{abstract}
%


\section{Introduction}
\label{sec:introduction}
\begin{figure}[t]
 \centering
\includegraphics[scale=0.9]{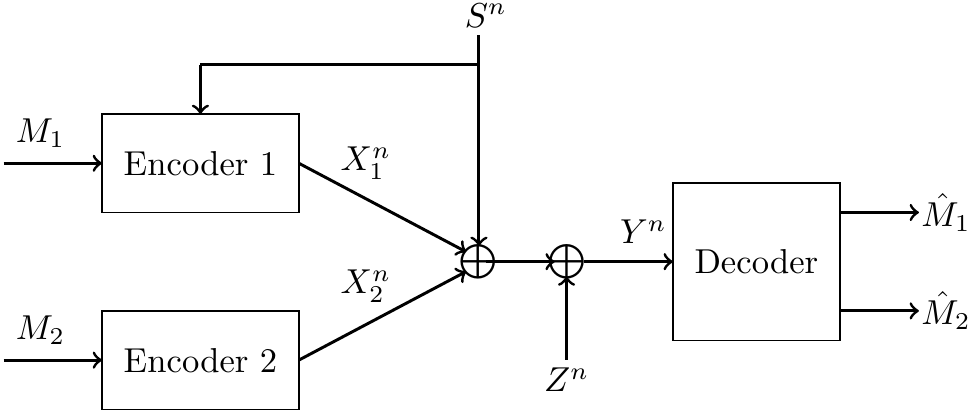}
\caption{State-dependent Gaussian MAC with state available noncausally at one encoder without degraded message sets. \label{fig:dirty-mac-setup}}
\end{figure}

\begin{figure}[t]
 \centering
\includegraphics[scale=0.9]{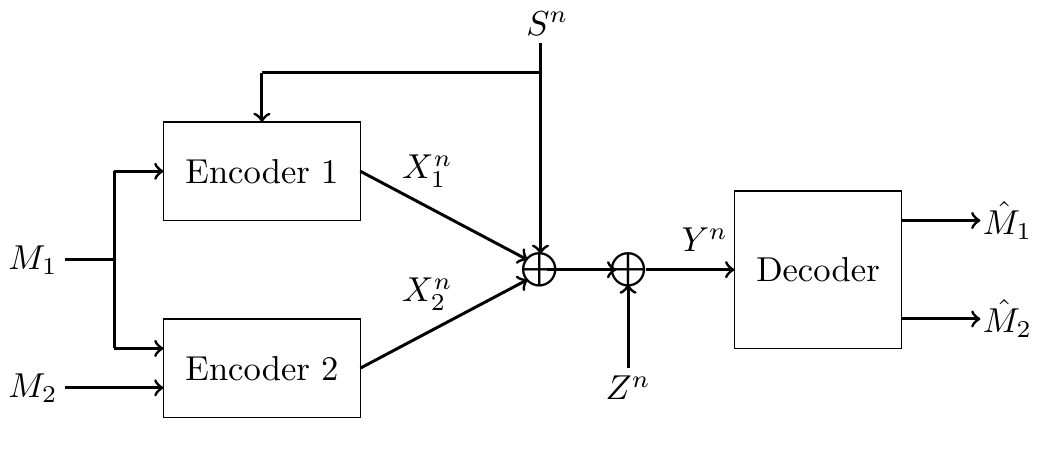}
\caption{State-dependent Gaussian MAC with state available noncausally at one encoder with degraded message sets. \label{fig:dirty-mac-setup-degraded}}
\end{figure}

We study a two-user state-dependent Gaussian multiple-access channel (MAC) with the state noncausally known at one encoder. 
The channel input-output relationship for a single channel use is given by 
\begin{IEEEeqnarray}{rCl}
Y = X_{1} + X_{2} + S + Z 
\label{eq:channel-io}
\end{IEEEeqnarray}
where  $Z \sim \mathcal{N}(0,1)$ denotes the additive white Gaussian noise, and $X_1$ and $X_2$ are the channel inputs from two users, which are subject to the (average) power constraints $P_1$ and $P_2$, respectively.
The state $S\sim \mathcal{N}(0,Q)$  is known noncausally at encoder 1 (state-cognitive user), but is not known at encoder~2 (non-cognitive user) nor at the decoder.
This channel model generalizes Costa's dirty-paper channel~\cite{costa1983-05a} to the multiple-access setting, and is  also known as  ``dirty MAC'' or ``MAC with a single dirty user''~\cite{philosof2011-08a}. 
In this paper, we consider the following two scenarios:
\begin{enumerate}[{i)}]
\item Each user wishes to communicate an \emph{independent} message to the common receiver, where the state-cognitive user sends the message $\msg_1$ and the non-cognitive user sends $\msg_2$ (see Fig.~\ref{fig:dirty-mac-setup});
\item The state-cognitive encoder sends the message $\msg_1$ and the non-cognitive encoder sends both $\msg_1$ and $\msg_2$ (see Fig.~\ref{fig:dirty-mac-setup-degraded}). In this case, the message $\msg_1$ can be also viewed as a common message. 
\end{enumerate}
We shall refer to the first setting as the ``dirty MAC without degraded message sets'', and the second setting as the ``dirty MAC with degraded message sets''.

Although the dirty MAC (with and without degraded message sets) described in~\eqref{eq:channel-io} has been studied extensively in the literature~\cite{philosof2011-08a,kotagiri2008-03a,philosof2009-06a,zaidi2009-06a}, no single-letter expression for the capacity region is characterized to date. 
For the dirty MAC without degraded message sets, Kotagiri and Laneman~\cite{kotagiri2008-03a} derived an inner bound on the capacity region using a generalized dirty paper coding scheme at the cognitive encoder, which allows arbitrary correlation between the input $X_1$ and the state $S$.
Philosof \emph{et al.}~\cite{philosof2011-08a} showed that the same rate region can be achieved by using lattice-based transmission.
In general, it is not clear whether a single-letter solution (i.e., random coding/random binning using independent and identically distributed (i.i.d.) copies of a certain scalar distribution) is optimal for the  dirty MAC~\eqref{eq:channel-io}.  
However, as~\cite{philosof2011-08a} and~\cite{philosof2009-06a} demonstrated, a single-letter solution is  suboptimal for the  \emph{doubly-dirty} MAC, in which the output is corrupted by two states, each known at one encoder noncausally (see also~\cite{pourbabaee2012-09a}). 
In this case, (linear) structured lattice coding outperforms the best known single-letter solution. 
An inner bound for the dirty MAC with degraded message sets was derived in~\cite{zaidi2009-06a}, which uses superposition coding at the non-cognitive encoder to send the two messages $\msg_1$ and $\msg_2$. 
On the converse side, all existing outer bounds for the dirty MAC without degraded message sets are obtained by assuming that a genie provides auxiliary information to the encoders/decoder.
For example, by revealing the state to the decoder, one obtains an  outer bound given by the capacity region of the Gaussian MAC without state dependence. 
In~\cite{zaidi2009-06a},  Zaidi \emph{et al.}  derived an outer bound on the capacity region of the dirty MAC with degraded message sets, which also serves as an outer bound for the dirty MAC without degraded message sets. 
Somekh-Baruch \emph{et al.}~\cite{somekh-baruch2008-10a} considered the setting in which the cognitive  encoder knows the message of the non-cognitive encoder (i.e.,  the roles of the two encoders are reversed), and derived the exact capacity region (see also~\cite{kotagiri2007-06a}). 
Interestingly, this capacity region remains valid if the non-cognitive encoder processes strictly causal state information~\cite{zaidi2013-10a}.

Different variants of the dirty MAC model in~\eqref{eq:channel-io} have also been investigated in the literature. 
A special case of the dirty MAC model is the ``helper problem''~\cite{mallik2008-01a}, in which the cognitive user does not send any information, and its goal is to help the non-cognitive user. For the helper problem, the capacity (of the non-cognitive user) is known for a wide range of channel parameters~\cite{sun2016-12a}. 
The authors in~\cite{lapidoth2010-06a} and~\cite{li2013-03a} considered the case in which the state is known only strictly causally or causally at the cognitive encoder, and derived inner and outer bounds on the capacity region. 
The capacity region of the MAC with action-dependent states was established in Dikstein \emph{et al.}~\cite{dikstein2015-01a}.
\ifthenelse{\boolean{conf}}{}
{
Finally, Wang~\cite{wang2012-05a} characterized the capacity region of the $K$-user dirty MAC to within a bounded gap.
For a general account of state-dependent multiuser models, we refer the reader to~\cite{jafar2006-12a} and~\cite{keshet2008-06a}.}

The main contributions of this paper are the establishment of new outer bounds on the capacity region of the dirty MAC given in~\eqref{eq:channel-io} with and without degraded message sets.
In both scenarios, our bounds improve uniformly over the best known outer bounds (see Fig.~\ref{fig:ca_region_full_non}--Fig.~\ref{fig:region-outer-degraded2} for numerical examples).
For the dirty MAC without degraded message sets, the new outer bounds allow us to characterize the two corner points of the capacity region as well as the sum rate capacity (note that, unlike~\cite{philosof2011-08a}, we do not assume $Q\to\infty$). 
In this case, a single-letter solution is shown to be adequate  to achieve both the corner points and the sum rate capacity. 
Furthermore, the full capacity region of the dirty MAC without degraded message sets is established in situations in which the sum rate capacity coincides with the capacity of the helper problem.

The proof of our outer bounds builds on a recent technique proposed by Polyanskiy and Wu~\cite{polyanskiy16-07a}  that  bounds the difference of the differential entropies of two probability distributions via their quadratic Wasserstein distance and via Talagrand's transportation inequality~\cite{talagrand1996-05a}.
It also relies on a generalized version of the  worst-case Gaussian noise result, in which the Gaussian input and the noise are dependent (but are uncorrelated)~\cite{ihara1978-04a,diggavi2001-11a,hassibi03}.
We anticipate  that these techniques can be useful more broadly for other state-dependent multiuser models, such as state-dependent interference channels and relay channels.

\section{Problem Setup and Previous Results}
\label{sec:problem-setup}

\subsection{Problem Setup}

Consider the Gaussian MAC~\eqref{eq:channel-io} with additive Gaussian state noncausally known at encoder~1 depicted in Fig.~\ref{fig:dirty-mac-setup} and Fig.~\ref{fig:dirty-mac-setup-degraded}.
The state $S~\sim \realg(0,Q)$ is independent of the additive white Gaussian noise $Z~\sim \realg(0,1)$ and of the input $X_2$ of the non-cognitive encoder. 
The state and the noise are i.i.d. over channel uses.
For the dirty MAC without degraded message sets (Fig.~\ref{fig:dirty-mac-setup}), we assume that encoder 1 and encoder 2 must satisfy the (average) power constraints\footnote{Note that, the authors of~\cite{philosof2011-08a} and~\cite{somekh-baruch2008-10a} assumed \emph{per-codeword} power constraints, i.e., for all messages $m_1$ and $m_2$,  the codewords  $x_1^n$ and $x_2^n$ satisfy $\sum\nolimits_{i=1}^{n} x_{1,i}^2 (m_1, S^n) \leq nP_1$ and $\sum\nolimits_{i=1}^{n} X_{2,i}^2(m_2) \leq nP_2$ almost surely. Clearly, every outer bound for the average power constraint is also a valid outer bound for the \emph{per-codeword} power constraint. }
\begin{IEEEeqnarray}{rCl}
\sum\limits_{i=1}^{n} \Ex{}{X_{1,i}^2(\msg_1, S^n)} &\leq& nP_1 \label{eq:power-constraint-1}\IEEEeqnarraynumspace \\
\sum\limits_{i=1}^{n} \Ex{}{X_{2,i}^2(\msg_2)} &\leq& nP_2  \label{eq:power-constraint-2}
\end{IEEEeqnarray}
where the index $i$ denotes the channel use, and $M_1$ and $M_2$ denote the transmitted messages, which are independently and uniformly distributed.
The decoder reconstructs the transmitted messages $\msg_1$ and $\msg_2$ from the channel output, and outputs $\hat{\msg}_1$ and $\hat{\msg}_2$. 
The (average) probability of error is defined as 
\begin{IEEEeqnarray}{rCl}
P_e\define \prob[(\msg_1,\msg_2) \neq (\hat{\msg}_1,\hat{\msg}_2 )].
\end{IEEEeqnarray}
If the message sets are degraded (Fig.~\ref{fig:dirty-mac-setup-degraded}), then the power constraint~\eqref{eq:power-constraint-2} becomes 
\begin{IEEEeqnarray}{rCl}
\sum\limits_{i=1}^{n} \Ex{}{X_{2,i}^2(\msg_1,\msg_2)} &\leq& nP_2.  \label{eq:power-constraint-2-degraded}
\end{IEEEeqnarray}

The capacity regions for the dirty MAC with and without degraded message sets are denoted by $\setC_{\deg}(P_1,P_2,Q)$ and $\setC(P_1,P_2,Q)$, respectively. Note that, by definition,
\begin{IEEEeqnarray}{rCl}
\setC(P_1,P_2,Q) \subseteq \setC_{\deg}(P_1,P_2,Q).
\end{IEEEeqnarray}
In both scenarios, a single-letter characterization for the capacity region is not known in the literature. In Section~\ref{sec:previous-results} below, we review the existing inner and outer bounds on $\setC_{\deg}(P_1,P_2,Q)$ and $\setC(P_1,P_2,Q)$.

\subsection{Previous Results}
\label{sec:previous-results}
For the dirty MAC without degraded message sets, the best known achievable rate region was derived by Kotagiri and Laneman~\cite{kotagiri2008-03a}, and is  given by the convex hull of the rate pairs $(R_1,R_2)$ satisfying 
\begin{IEEEeqnarray}{rCl}
R_1 &\leq& I(U;Y|X_2) -I(U;S) \label{eq:inner-bound-kotagiri-1}\\
R_2 &\leq& I(X_2;Y|U)\\
R_1+R_2 &\leq& I(U,X_2;Y) - I(U,S) \label{eq:inner-bound-kotagiri-3}
\end{IEEEeqnarray}
for some joint probability distribution $P_{UX_1|S}P_{X_2}$. 
A computable inner bound was obtained in~\cite{kotagiri2008-03a} from~\eqref{eq:inner-bound-kotagiri-1}--\eqref{eq:inner-bound-kotagiri-3} by setting  
\begin{IEEEeqnarray}{rCl}
P_{X_1|S=s}&=& \mathcal{N}\lefto(\rho \sqrt{P_1/Q}s, P_1(1-\rho^2)\right) \label{eq:generalized-dpc-1}\\
 P_{X_2} &= & \mathcal{N}(0,P_2)\label{eq:generalized-dpc-2}\\
U &=& X_1-\rho\sqrt{\frac{P_1}{Q}}S + \alpha\left(1+\rho\sqrt{\frac{P_1}{Q}}\right)S  \IEEEeqnarraynumspace \label{eq:generalized-dpc-3}
\end{IEEEeqnarray}
for some $\rho\in[-1,0]$ and $\alpha \in \realset$.
This choice of input distribution is also known as \emph{generalized dirty paper coding}.
Unlike in the point-to-point setting~\cite{costa1983-05a}, allowing a (negative) correlation between $X_1$ and $S$ may be beneficial  since it partially cancels  the state for the non-cognitive encoder. 
However, it is not clear whether the Gaussian distribution optimizes the bounds in~\eqref{eq:inner-bound-kotagiri-1}--\eqref{eq:inner-bound-kotagiri-3}. 

%
%

The best known outer bound is  given by the region of rate pairs $(R_1,R_2)$ satisfying\footnote{In this paper, the logarithm ($\log$) and exponential ($\exp$) functions are taken with respect to an arbitrary basis.} 
\begin{IEEEeqnarray}{rCl}
\qquad\,\, R_1 &\leq &\frac{1}{2}\log(1+P_1(1-\rho_1^2-\rho_s^2)) \label{eq:outer-somekh1}\qquad\qquad\qquad\qquad \\
\qquad\,\, R_2 &\leq& \frac{1}{2}\log\left(1+ \frac{P_2(1-\rho_1^2-\rho_s^2)}{1-\rho_s^2}\right)\label{eq:outer-zaidi}\\
R_1 +R_2 &\leq& \frac{1}{2}\log(1+P_1(1-\rho_1^2-\rho_s^2))  \qquad\qquad\qquad\qquad\notag\\
&&+\,\frac{1}{2} \log\mathopen{}\Big(1+ \frac{(\sqrt{P_2} + \rho_1 \sqrt{P_1})^2}{1+P_1(1-\rho_1^2-\rho_s^2) +(\sqrt{Q} +\rho_s\sqrt{P_1})^2 }\Big)\IEEEeqnarraynumspace \label{eq:outer-somekh2}\\
R_1+R_2 &\leq& \frac{1}{2}\log(1+P_1+P_2)
\label{eq:ob-sum-rate-mac}
\end{IEEEeqnarray}
for some $\rho_1\in[0,1]$ and $\rho_s\in [-1,0]$ that satisfy  $\rho_1^2 +\rho_s^2 \leq 1$.
This outer bound is a combination of  several (genie-aided) outer bounds established in the literature:
\begin{itemize}
\item The bounds~\eqref{eq:outer-zaidi} and~\eqref{eq:outer-somekh2} form the outer bound in~\cite{zaidi2009-06a} on $\setC_{\deg}(P_1,P_2,Q)$, and hence on $\setC(P_1,P_2,Q)$.
\item The bounds~\eqref{eq:outer-somekh1} and~\eqref{eq:outer-somekh2} characterize the capacity region of the dirty MAC under the assumption  that the cognitive user knows the message of the non-cognitive user~\cite{somekh-baruch2008-10a}.
\item The bound~\eqref{eq:ob-sum-rate-mac} upper-bounds the sum rate  of the Gaussian MAC without state dependence. 
\end{itemize} 

For the dirty MAC with degraded message sets,  inner and outer bounds  on the capacity region were derived  in~\cite{zaidi2009-06a}. 
As reviewed above, the capacity region $\setC_{\deg}(P_1,P_2,Q)$ is outer-bounded by the region with rate pairs $(R_1,R_2)$ satisfying~\eqref{eq:outer-zaidi} and~\eqref{eq:outer-somekh2}. 
This outer bound follows from the following single-letter outer region~\cite[Th.~2]{zaidi2009-06a}:
\begin{IEEEeqnarray}{rCl}
R_2&\leq& I(X_2;Y|S,X_1) \label{eq:zaidi-09-rate2}\\
R_1+R_2&\leq& I(X_1,X_2;Y|S) - I(S;X_2|Y) \label{eq:zaidi-09-sum-rate}
\end{IEEEeqnarray}
where the joint probability distributions of $X_1$, $X_2$, and $S$ must be of the form $P_{S}P_{X_2} P_{X_1|X_2,S}$.
The inner bound in~\cite{zaidi2009-06a} consists of rate pairs $(R_1,R_2)$ satisfying
\begin{IEEEeqnarray}{rCl}
R_2 &\leq& I(X_2;Y|U_1,U_2)\\
R_2 &\leq& I(X_2,U_2;Y|U_1) - I(U_2;S|U_1)\\
R_1+R_2 &\leq&  I(X_2,U_1,U_2;Y) - I(U_2;S|U_1)
\end{IEEEeqnarray}
for some joint probability distributions $P_SP_{U_1}P_{X_2|U_1}P_{U_2|U_1,S}P_{X_1|U_1,U_2,S}$ that satisfy 
\begin{IEEEeqnarray}{rCl}
I(U_2;Y|U_1,X_1) - I(U_2;S|U_1) \geq 0. 
\end{IEEEeqnarray}
This inner bound is evaluated in~\cite{zaidi2009-06a} for the case in which $(X_1,X_2,U_1,U_2,S)$ are jointly Gaussian distributed. 
Again,  it is not known whether the Gaussian input optimizes the  bound.

\subsection{The Helper Problem}

As reviewed in the introduction, the dirty MAC model includes the helper problem as a special case. More specifically, in the helper problem, the cognitive user (also known as the \emph{helper}) does not send any information, and its goal is to assist the non-cognitive user by canceling the state.
The capacity of the helper problem is defined as 
\begin{IEEEeqnarray}{rCl}
C_{\mathrm{helper}} &\define& \max\{R_2: (0,R_2) \in \setC(P_1,P_2,Q)\} \label{eq:def-helper-capacity}\\
&=&  \max\{R_2: (0,R_2) \in \setC_{\deg}(P_1,P_2,Q)\}  . \label{eq:def-helper-capacity-def2}
\end{IEEEeqnarray}
The equivalence between~\eqref{eq:def-helper-capacity} and~\eqref{eq:def-helper-capacity-def2} follows  since  $I(M_1;X_2^n) = 0$ regardless of whether the message sets are degraded or not.

The capacity of the helper problem was studied in~\cite{mallik2008-01a} and~\cite{sun2016-12a}, and is known for a wide range of channel parameters. 
More specifically, it was shown that~\cite[Th.~2]{sun2016-12a}
\begin{IEEEeqnarray}{rCl}
C_{\mathrm{helper}} = \frac{1}{2}\log(1+P_2)
\end{IEEEeqnarray}
provided that $P_1$, $P_2$, and $Q$ satisfy the following condition. 

\begin{cond}
\label{cond:parameters}
There exists an $\alpha \in [1-\sqrt{P_1/Q}, 1+ \sqrt{P_1/Q}]$ such that
\begin{IEEEeqnarray}{rCl}
(P_1- (\alpha-1)^2Q)^2 \geq \alpha^2 Q(P_2+1 - P_1+ (\alpha-1)^2 Q). \IEEEeqnarraynumspace
\label{eq:cond-1}
\end{IEEEeqnarray}
\end{cond}

In other words, if Condition~\ref{cond:parameters} is satisfied, then the state can be perfectly canceled, and the non-cognitive user achieves the channel capacity without state dependence.  Note that, to satisfy Condition~\ref{cond:parameters} it is not necessary that $P_1 \geq Q$ (e.g.,  \eqref{eq:cond-1}  holds as long as  $P_1 \geq P_2 +1$, regardless of the value of~$Q$).

\section{Main Results}
\label{sec:main-result}

The main results of this paper are the establishment of several new outer bounds on the capacity region of the dirty MAC~\eqref{eq:channel-io} with and without degraded message sets. 
%
%
%
%
%
%
For notational convenience, we denote 
\begin{IEEEeqnarray}{rCl}
C_1 \define \frac{1}{2}\log(1+P_1) ,\quad C_2 \define \frac{1}{2}\log(1+P_2).
\end{IEEEeqnarray}

\subsection{Dirty MAC Without Degraded Message Sets}
\label{sec:outer-bounds-mac}

\subsubsection{New outer bounds} In this section, we present two outer bounds on  $\setC(P_1,P_2,Q)$.

\begin{thm}
\label{thm:outer-bound-mac}
%
The capacity region $\setC(P_1,P_2,Q)$ of the dirty MAC without degraded message sets is outer-bounded by the region with rate pairs $(R_1,R_2)$ satisfying
\begin{IEEEeqnarray}{rCl}
R_2 \leq C_{\mathrm{helper}}
\label{eq:ub-r2-trivial}
\end{IEEEeqnarray}
and
\begin{IEEEeqnarray}{rCl}
R_1\leq \min_{0\leq \delta \leq 1}  \Big\{ \frac{1}{2} \log\lefto( 1 + \frac{1+P_2-\delta}{ P_2\delta} g(R_2)\right) + f(\delta) \Big\} \IEEEeqnarraynumspace
\label{eq:thm1-ub}
\end{IEEEeqnarray}
where
\begin{IEEEeqnarray}{rCl}
g(R_2) \define \exp\lefto(2 c_1\sqrt{C_2-R_2} + 2(C_2-R_2)\right) -1 \IEEEeqnarraynumspace
\label{eq:def-g-r2}
\end{IEEEeqnarray}
with
\begin{IEEEeqnarray}{rCl}
c_1\define \frac{3\sqrt{1+(\sqrt{P_1}+\sqrt{Q})^2+P_2} + 4(\sqrt{P_1} + \sqrt{Q}) }{\sqrt{(1+P_2)/(2\log e)}} \IEEEeqnarraynumspace
\label{eq:def-c1}
\end{IEEEeqnarray}
and
\ifthenelse{\boolean{singcol}}{
\begin{IEEEeqnarray}{rCl}
f(\delta) &\define& \max_{\rho\in[-1,0]}  \frac{1}{2} \Big\{ \log  \frac{1+P_2 + P_1+Q+  2\rho\sqrt{P_1Q} }{\delta + P_1+Q+  2\rho\sqrt{P_1Q}}+ \log \frac{ \delta + (1-\rho^2) P_1}{1+P_2} \Big\}.\IEEEeqnarraynumspace
\label{eq:def-f-delta}
\end{IEEEeqnarray}
}{
\begin{IEEEeqnarray}{rCl}
f(\delta) &\define& \max_{\rho\in[-1,0]}  \frac{1}{2} \Big\{ \log  \frac{1+P_2 + P_1+Q+  2\rho\sqrt{P_1Q} }{\delta + P_1+Q+  2\rho\sqrt{P_1Q}}\notag\\
&&\qquad\qquad\qquad\qquad\quad \,\,\,+\,\log \frac{ \delta + (1-\rho^2) P_1}{1+P_2} \Big\}.\IEEEeqnarraynumspace
\label{eq:def-f-delta}
\end{IEEEeqnarray} }
\end{thm}
\begin{IEEEproof}
See Section~\ref{sec:proof-thm-polywu}.
\end{IEEEproof}
\begin{rem}
\label{rem:concave-f}
The objective function on the right-hand side (RHS) of~\eqref{eq:def-f-delta} is concave in $\rho$ for every $\delta\in[0,1]$.
\end{rem}
\begin{rem}
The upper bound~\eqref{eq:thm1-ub} can be slightly improved by replacing $Q$ on the RHS of~\eqref{eq:thm1-ub} with $\widetilde{Q}\leq Q$ and by minimizing over $\widetilde{Q}$.  This follows because, for a fixed rate $R_2$, the maximum achievable $R_1$ is monotonically non-increasing in $Q$, whereas the RHS of~\eqref{eq:thm1-ub} is not. 
\end{rem}


We next illustrate the main intuition behind Theorem~\ref{thm:outer-bound-mac}.
To concentrate ideas, we assume that the channel parameters $P_1$, $P_2$, and $Q$ satisfy Condition~\ref{cond:parameters}, which implies that $C_{\mathrm{helper}} = C_2$~\cite[Th.~2]{sun2016-12a}. 
Consider two auxiliary channels 
\begin{IEEEeqnarray}{rCl}
Y_G^n &\define& X_1^n + S^n + G^n + Z^n\label{eq:def-ydelta-n}\\
Y_\delta^n &\define& X_1^n + S^n + \sqrt{\delta} Z^n
\label{eq:def-yg-n}
\end{IEEEeqnarray}
where $G^n~\sim \realg(0, P_2\matI_n)$ is a Gaussian vector having the same power as $X_2^n$, and $\delta \in (0,1)$ is a constant. 
In words, $Y_G^n$ is obtained from $Y^n$ by replacing the codeword $X_2^n$ with Gaussian interference of the same power, 
and $Y_\delta^n$ is obtained from $Y^n$ by removing the interference $X^n_2$ and by increasing the signal-to-noise ratio (SNR).
Therefore, the channel $M_1 \to {Y_G^n}$ is worse than the original channel whereas the channel $M_1 \to Y_{\delta}^n$ is better than the original one.
In fact, we argue next that, when the non-cognitive user is communicating at a rate close to its maximum rate $C_2$, the three channels have approximately the same rate for the cognitive user.

Indeed, suppose that $R_2 \approx C_2$. 
Then, on the one hand, the distribution of $X_2^n$ is close to that of $G^n$, and hence 
 \begin{IEEEeqnarray}{rCl}
I(X^n_1+S^n;Y_G^n) \approx  I(X^n_1+S^n;Y^n).
\label{eq:intuition-y-g}
\end{IEEEeqnarray}
 On the other hand, since the receiver is able to decode the message of the non-cognitive user, it follows that 
 \begin{IEEEeqnarray}{rCl}
 I(X^n_1+S^n;Y^n) &\approx& I(X^n_1+S^n;Y^n |X_2^n)\\
 & =& I(X^n_1+S^n;X^n_1+S^n +Z^n). \IEEEeqnarraynumspace
 \label{eq:intuition-y-1}
\end{IEEEeqnarray}
Combining~\eqref{eq:intuition-y-g} and~\eqref{eq:intuition-y-1}, we conclude that 
\begin{IEEEeqnarray}{rCl}
\IEEEeqnarraymulticol{3}{l}{
I(X^n_1+S^n;X_1^n+S^n + G^n +Z^n) }\notag \\
& \approx& I(X^n_1+S^n;X^n_1+S^n +Z^n).
\end{IEEEeqnarray}
In other words, reducing the power of the Gaussian noise (from $1+P_2$ to $1$) does not (significantly) increase the mutual information between $X_1^n+S^n$ and the output.  
By further reducing the noise power, we obtain   
\begin{equation} 
 I(X^n_1+S^n;Y^n) \approx I(X^n_1+S^n;Y_G^n)  \approx I(X^n_1+S^n;Y^n_\delta).
\label{eq:intuition-g-equal-delta}
\end{equation}
The errors in the estimation~\eqref{eq:intuition-g-equal-delta} can be bounded via Costa's entropy power inequality~\cite{costa1985-11a} or the I-MMSE relation~\cite{guo2005-04a}.

To see how the relation~\eqref{eq:intuition-g-equal-delta} can be used to upper-bound $R_1$,  we note that by standard manipulations of mutual information,
\begin{IEEEeqnarray}{rCl}
nR_1 &\leq& I(X_1^n + S^n; Y^n) - I(S^n;Y^n).  \label{eq:intuition-r1}
\end{IEEEeqnarray}
By~\eqref{eq:intuition-g-equal-delta}, we may replace the two $Y^n$'s on the RHS of~\eqref{eq:intuition-r1} with $Y_G^n$ and $Y_\delta^n$, respectively, and  obtain
\begin{IEEEeqnarray}{rCl}
nR_1 &\lessapprox&  I(X_1^n + S^n; Y^n_G) - I(S^n;Y^n_\delta)\\
&\lessapprox&  n \max_{P_{X_1|S}} \Big\{ I(X_1+S; Y_G) - I(S; Y_{\delta}) \Big\}
\label{eq:intuitive-max-gaussian}
\end{IEEEeqnarray}
where
\begin{IEEEeqnarray}{rCl}
Y_G &\define& X_1+S+ G+ Z\label{eq:def-yg-sing}\\
Y_\delta &=& X_1+S+\sqrt{\delta} Z\label{eq:def-ydelta-sing}
\end{IEEEeqnarray}
are the single-letter versions of $Y_G^n$ and $Y_\delta^n$, respectively.
By the Gaussian saddle point property (namely, the Gaussian distribution is the best input distribution for Gaussian noise, and is the worst noise distribution for a Gaussian input), we expect that the RHS of~\eqref{eq:intuitive-max-gaussian} is maximized when $(X_1,S)$ are jointly Gaussian. 
The maximum of the objective function on the RHS of~\eqref{eq:intuitive-max-gaussian} is precisely the $f(\delta)$ defined in~\eqref{eq:def-f-delta},
whereas the logarithm term on the RHS of~\eqref{eq:thm1-ub} quantifies the error in the approximation~\eqref{eq:intuition-g-equal-delta}, which vanishes as $R_2\to C_2$.
The rigorous proof of Theorem~\ref{thm:outer-bound-mac} which builds upon the above intuition can be found in Section~\ref{sec:proof-thm-polywu}.

 %

 The outer bound provided in Theorem~\ref{thm:outer-bound-mac} improves the best known outer bound in the regime where $R_2$ is close to $C_2$ (provided that $C_{\mathrm{helper}}$ is also close to $C_2$). 
 The next theorem provides a tighter upper bound on the sum rate than~\eqref{eq:outer-somekh2} and~\eqref{eq:ob-sum-rate-mac}.

%

\begin{thm}
\label{thm:outer-bound-r2}
The capacity region $\setC(P_1,P_2,Q)$ of the dirty MAC without degraded message sets is outer-bounded by the region with rate pairs $(R_1,R_2)$ satisfying
\begin{IEEEeqnarray}{rCl}
R_1&\leq& \frac{1}{2}\log(1+P_1(1-\rho^2)) \label{eq:rate-r2-outer-bound1}\\
R_2 &\leq& C_{\mathrm{2}}\label{eq:rate-r2-outer-bound2}\\
R_1+R_2 &\leq& \frac{1}{2}\log\lefto(1+\frac{P_2}{1+P_1+Q+2\rho\sqrt{P_1Q}}\right) \notag\\
&&+\,\frac{1}{2}\log(1+P_1(1-\rho^2))
\label{eq:rate-r2-outer-bound}
\end{IEEEeqnarray}
for some $\rho \in [-1,0]$.
\end{thm}

 \begin{IEEEproof}
The proof of Theorem~\ref{thm:outer-bound-r2} follows from the following single-letter outer bound on the capacity region. 
\begin{prop}
\label{prop:outer-bound-general}
The capacity region $\setC(P_1,P_2,Q)$ of the dirty MAC without degraded message sets is outer-bounded by the region with rate pairs $(R_1,R_2)$ satisfying
\begin{IEEEeqnarray}{rCl}
R_1&\leq& I(X_1;Y|X_2,S) \label{eq:rate-r2-outer-bound1-general}\\
R_2 &\leq& I(X_2;Y|X_1,S)\label{eq:rate-r2-outer-bound2-general}\\
R_1+R_2 &\leq& I(X_1;Y|X_2,S) + I(X_2;Y) 
\label{eq:rate-r2-outer-bound-general}
\end{IEEEeqnarray}
for some joint distributions $P_SP_{X_1|S} P_{X_2}$ that satisfy the power constraint 
\begin{IEEEeqnarray}{rCl}
\Ex{}{X_1^2} \leq P_1\text{  and  } \Ex{}{X_2^2} \leq P_2. 
\end{IEEEeqnarray}
\end{prop}
\begin{IEEEproof}
See Section~\ref{sec:proof-prop-outer-nondegraded}.
\end{IEEEproof}
It is not difficult to show that the outer bound in Proposition~\ref{prop:outer-bound-general} is maximized when $S$, $X_1$, and $X_2$ are jointly Gaussian distributed (proof omited). 
Evaluating~\eqref{eq:rate-r2-outer-bound1-general}--\eqref{eq:rate-r2-outer-bound-general} for Gaussian distributions $P_SP_{X_1|S} P_{X_2}$, we obtain the outer bound in Theorem~\ref{thm:outer-bound-r2}.
 \end{IEEEproof}

\subsubsection{Sum rate capacity}
\label{sec:sum-rate-capacity}
Let $C_{\mathrm{sum}}$ be the sum rate capacity of the dirty MAC~\eqref{eq:channel-io} without degraded message sets, i.e., 
\begin{IEEEeqnarray}{rCl}
C_{\mathrm{sum}} \define \max\{ R_1+R_2: (R_1,R_2) \in \setC(P_1,P_2,Q) \}. \IEEEeqnarraynumspace
\end{IEEEeqnarray}
By comparing the inner bound~\eqref{eq:inner-bound-kotagiri-3} (evaluated using Gaussian inputs) and the outer bound~\eqref{eq:rate-r2-outer-bound}, we establish  the sum rate capacity $C_{\mathrm{sum}}$.

\begin{thm}
\label{thm:sum-rate-capacity}
The sum rate capacity of the dirty MAC without degraded message  sets~is  given by
\ifthenelse{\boolean{singcol}}{
\begin{IEEEeqnarray}{rCl}
C_{\mathrm{sum}} =  \max_{\rho \in[-1,0]} \,&&\frac{1}{2}\Big\{ \log\mathopen{}\Big(1+\frac{P_2}{1+P_1+Q+2\rho\sqrt{P_1Q}}\Big)  + \frac{1}{2}\log(1+P_1(1-\rho^2))\Big\}\IEEEeqnarraynumspace \label{eq:sum-rate-capacity}
\end{IEEEeqnarray}
}{
\begin{IEEEeqnarray}{rCl}
C_{\mathrm{sum}} =  \max_{\rho \in[-1,0]} \,&&\frac{1}{2}\Big\{ \log\mathopen{}\Big(1+\frac{P_2}{1+P_1+Q+2\rho\sqrt{P_1Q}}\Big) \notag\\
&&\qquad\qquad\quad +\,\frac{1}{2}\log(1+P_1(1-\rho^2))\Big\}\IEEEeqnarraynumspace \label{eq:sum-rate-capacity}
\end{IEEEeqnarray}}
or equivalently, 
\begin{IEEEeqnarray}{rCl}
C_{\mathrm{sum}} = C_2 + f(1) .
\label{eq:sum-rate-capacity-alt}
\end{IEEEeqnarray}
\end{thm}

\begin{IEEEproof}
The converse part of~\eqref{eq:sum-rate-capacity} follows directly from~\eqref{eq:rate-r2-outer-bound}. 
Since the objective function on the RHS of~\eqref{eq:sum-rate-capacity} is continuous and concave in $\rho\in [-1,0]$ (see Remark~\ref{rem:concave-f}), it has a unique maximizer on $[-1,0]$, which we denote by 
$\rho^*$. 
It follows that the rate pair 
\begin{IEEEeqnarray}{rCl}
\bar{R}_1 &\define& \frac{1}{2}\log(1+P_1(1-(\rho^*)^2))\label{eq:sum-rate-ach1}\\
\bar{R}_2 &\define &  \frac{1}{2}\log\lefto(1+\frac{P_2}{1+P_1+Q+2\rho^*\sqrt{P_1Q}}\right) \label{eq:sum-rate-ach2}
\end{IEEEeqnarray}
is achievable by treating the interference $X_1+S$ as noise for the non-cognitive user, and by using generalized dirty paper coding for the cognitive user  with $\rho =\rho^*$ and 
\begin{IEEEeqnarray}{rCl}
\alpha = \frac{P_1(1-(\rho^*)^2)}{P_1(1-(\rho^*)^2)+1}
\label{eq:alpha-choice}
\end{IEEEeqnarray}
in~\eqref{eq:generalized-dpc-1}--\eqref{eq:generalized-dpc-3}.
The choice of $\alpha$ in~\eqref{eq:alpha-choice} is the usual dirty paper coding coefficient for the equivalent channel (obtained by canceling the interference $X_2$ from the non-cognitive user)
\begin{IEEEeqnarray}{rCl}
\widetilde{Y} = X_0 + \left(1-\rho^*\sqrt{\frac{P_1}{Q}}\right)S + Z
\end{IEEEeqnarray}
where $X_0 \define X_1 - \rho^*\sqrt{P_1/Q} S \sim \mathcal{N}(0,P_1(1-(\rho^*)^2))$ is independent of $S$. 
The rate pair in~\eqref{eq:sum-rate-ach1} and~\eqref{eq:sum-rate-ach2} achieves the sum rate capacity~\eqref{eq:sum-rate-capacity}. %
The equivalence between~\eqref{eq:sum-rate-capacity} and~\eqref{eq:sum-rate-capacity-alt} is straightforward to establish. 
\end{IEEEproof}

The next result shows that, if $C_{\mathrm{helper}} =C_{\mathrm{sum}}$, then the outer bound in Theorem~\ref{thm:outer-bound-r2} matches the
\ifthenelse{\boolean{conf}}{inner bound in~\cite{kotagiri2008-03a}.}{inner bound  in~\eqref{eq:inner-bound-kotagiri-1}--\eqref{eq:inner-bound-kotagiri-3} evaluated for Gaussian inputs.}
In this case, we obtain a complete characterization of the capacity region $\setC(P_1,P_2,Q)$.

\begin{cor}
\label{cor:complete-characterization-1}
For the dirty MAC without degraded messages, if $C_{\mathrm{helper}} =C_{\mathrm{sum}}$, then  the capacity region is given by the  convex hull of the set of rate pairs $(R_1,R_2)$ satisfying   
\begin{IEEEeqnarray}{rCl}
R_1&\leq& \frac{1}{2}\log\mathopen{}\big(1+P_1(1-\rho^2)\big) \label{eq:rate-r2-outer-bound1-special}\\
R_1+R_2 &\leq& \frac{1}{2}\log\lefto(1+\frac{P_2}{1+P_1+Q+2\rho\sqrt{P_1Q}}\right) \notag\\
&&+\,\frac{1}{2}\log(1+P_1(1-\rho^2))
\label{eq:rate-r2-outer-bound-special}
\end{IEEEeqnarray}
for some $\rho \in [-1,0]$.
\end{cor}
\begin{IEEEproof}
By Theorem~\ref{thm:outer-bound-r2}, the rate region characterized by~\eqref{eq:rate-r2-outer-bound1-special} and~\eqref{eq:rate-r2-outer-bound-special}, which we denote by $\setR^*(P_1,P_2,Q)$, is an outer bound on the capacity region $\setC(P_1,P_2,Q)$. 

To prove Corollary~\ref{cor:complete-characterization-1}, it suffices to show that the rate region $\setR^*(P_1,P_2,Q)$ is achievable.
Observe that, by the hypothesis $C_{\mathrm{helper}} = C_{\mathrm{sum}}$, the sum rate capacity is achieved with the rate pairs $(0,C_{\mathrm{helper}})$ and $(\bar{R}_1, \bar{R}_2)$, where $\bar{R}_1$ and $\bar{R}_2$ are defined in~\eqref{eq:sum-rate-ach1} and~\eqref{eq:sum-rate-ach2}, respectively.
Let now $(R_1,R_2)$ be an arbitrary point that lies on the boundary of $\setR^*(P_1,P_2,Q)$. If $R_1\leq \bar{R}_1$, then the rate pair $(R_1,C_{\mathrm{sum}}-R_1)$  is achievable using time sharing.
Since, by~\eqref{eq:rate-r2-outer-bound-special}, $R_2 \leq C_{\mathrm{sum}} -R_1$, we conclude that the rate pair $(R_1,C_{\mathrm{sum}} -R_1)$ coincides with $(R_1,R_2)$. 
If $\bar{R}_1 \leq R_1 \leq C_1$, it follows that there exists an $\rho_0 \in[\rho^*, 0]$ which satisfies $R_1 = \frac{1}{2}\log(1+P_1(1-\rho_0^2))$.
In this case, we have 
\begin{IEEEeqnarray}{rCl}
R_2= \frac{1}{2}\log\lefto(1+\frac{P_2}{1+P_1+Q+2\rho_0\sqrt{P_1Q}}\right).
\end{IEEEeqnarray}
This rate pair is again achievable by treating interference as noise for the non-cognitive user, and by using generalized dirty paper coding for the cognitive user. 
\end{IEEEproof}

For the case when $C_{\mathrm{helper}} < C_{\mathrm{sum}}$, the outer bound in Theorem~\ref{thm:outer-bound-r2} matches the inner bound only for $R_1$ values greater than a threshold $R_{1,\mathrm{th}}$. This threshold is given by
\begin{IEEEeqnarray}{rCl}
R_{1,\mathrm{th}} = I(U^*;Y)-I(U^*;S)
\label{eq:R_1_th_def}
\end{IEEEeqnarray}  
where $X_1^*$, $X_2^*$, and $U^*$ are given in~\eqref{eq:generalized-dpc-1}--\eqref{eq:generalized-dpc-3} with $\rho$ and $\alpha$ chosen as in the proof of Theorem~\ref{thm:sum-rate-capacity}.   
It is also not difficult to check that  $R_{1,\mathrm{th}} > 0$   if and only if  $C_{\mathrm{helper}} < C_{\mathrm{sum}}$.

\subsubsection{Corner points}
\label{sec:corner-points}
The bounds in Theorems~\ref{thm:outer-bound-mac} and~\ref{thm:outer-bound-r2} allow us to characterize the corner points  of the capacity region, which are defined as  
\begin{IEEEeqnarray}{rCl}
\widetilde{C}_1(P_1,P_2,Q) &\define& \max \{ R_1: (R_1, C_2) \in \setC(P_1,P_2,Q) \} \label{eq:corner-point-r1} \IEEEeqnarraynumspace\\
\widetilde{C}_2(P_1,P_2,Q) &\define& \max \{ R_2: (C_1, R_2) \in \setC(P_1,P_2,Q) \}. \IEEEeqnarraynumspace
\end{IEEEeqnarray}

\begin{cor}
For every $P_1$, every $P_2$, and every $Q$, we have 
\begin{IEEEeqnarray}{rCl}
\widetilde{C}_2(P_1,P_2,Q) = \frac{1}{2}\log\lefto(1+ \frac{P_2}{1+P_1+Q}\right). 
\label{eq:corner-point-bottom}
\end{IEEEeqnarray}
Furthermore, if $P_1$, $P_2$, and $Q$ satisfy Condition~\ref{cond:parameters},  then
\begin{IEEEeqnarray}{rCl}
\widetilde{C}_1(P_1,P_2,Q) = f(0) 
\label{eq:corner-point-top}
\end{IEEEeqnarray}
where $f(\cdot)$ is defined in~\eqref{eq:def-f-delta}.
\end{cor}
\begin{IEEEproof}
The corner point~\eqref{eq:corner-point-bottom} follows from~\eqref{eq:rate-r2-outer-bound1} and \eqref{eq:rate-r2-outer-bound} (with $\rho=0$), and~\eqref{eq:corner-point-top} follows from~\eqref{eq:thm1-ub} by setting $R_2=C_2$, and by taking $\delta\to 0$.
\end{IEEEproof}

A few remarks are in order.
\begin{itemize}
\ifthenelse{\boolean{conf}}{
\item The bottom corner point $(C_1, \widetilde{C}_2)$ also follows from the (genie-aided) outer bound in~\eqref{eq:outer-somekh1} and~\eqref{eq:outer-somekh2}.}{
\item The bottom corner point $(C_1, \widetilde{C}_2)$ also follows from the (genie-aided) outer bound~\eqref{eq:outer-somekh1} and \eqref{eq:outer-somekh2} developed in~\cite{somekh-baruch2008-10a}.
}
\item In the asymptotic limit of strong state power (i.e., $Q\to \infty$),  the two corner points become
\begin{IEEEeqnarray}{rCl}
\lim\limits_{Q\to \infty} \widetilde{C}_1(P_1,P_2,Q) &=& \frac{1}{2}\log \frac{P_1}{1+P_2} \\
\lim\limits_{Q\to \infty} \widetilde{C}_2(P_1,P_2,Q) &=& 0.
\end{IEEEeqnarray}
For comparison, existing outer bounds in~\cite{philosof2011-08a} and~\cite{zaidi2009-06a} only yield the upper bound 
\begin{IEEEeqnarray}{rCl}
\lim\limits_{Q\to \infty} \widetilde{C}_1(P_1,P_2,Q) &\leq& \frac{1}{2}\log \frac{1+P_1}{1+P_2}.  \IEEEeqnarraynumspace
\end{IEEEeqnarray}
\item  The top corner point $(\widetilde{C}_1, C_2)$ is achieved by using generalized dirty paper coding with $U=X_1+S$ and by treating the interference $X_2$ as noise for the cognitive user. The proof of Theorem~\ref{thm:outer-bound-mac} suggests that there is essentially no other alternative. Indeed, if $R_2 = C_2 + \littleo(1)$ as $n\to\infty$, then by~\eqref{eq:intuition-g-equal-delta} and the I-MMSE relation~\cite{guo2005-04a}, the minimum mean-square error (MMSE) in estimating $X_1^n+S^n$ given $Y_G^n$ satisfies 
\begin{IEEEeqnarray}{rCl}
\mathsf{MMSE}(X_1^n+S^n|Y_G^n) =\littleo(n).
\end{IEEEeqnarray}
This implies that,  in order to achieve $R_2= C_2+o(1)$, it is necessary for the decoder  to ``decode'' $X_1^n +S^n$ without knowing the codebook of the non-cognitive user (recall that $Y_G^n$ is obtained from $Y^n$ by replacing the codeword $X_2^n$ with Gaussian interference of the same power).
%
 %
\end{itemize}

\begin{figure}
\centering
\includegraphics[scale=1]{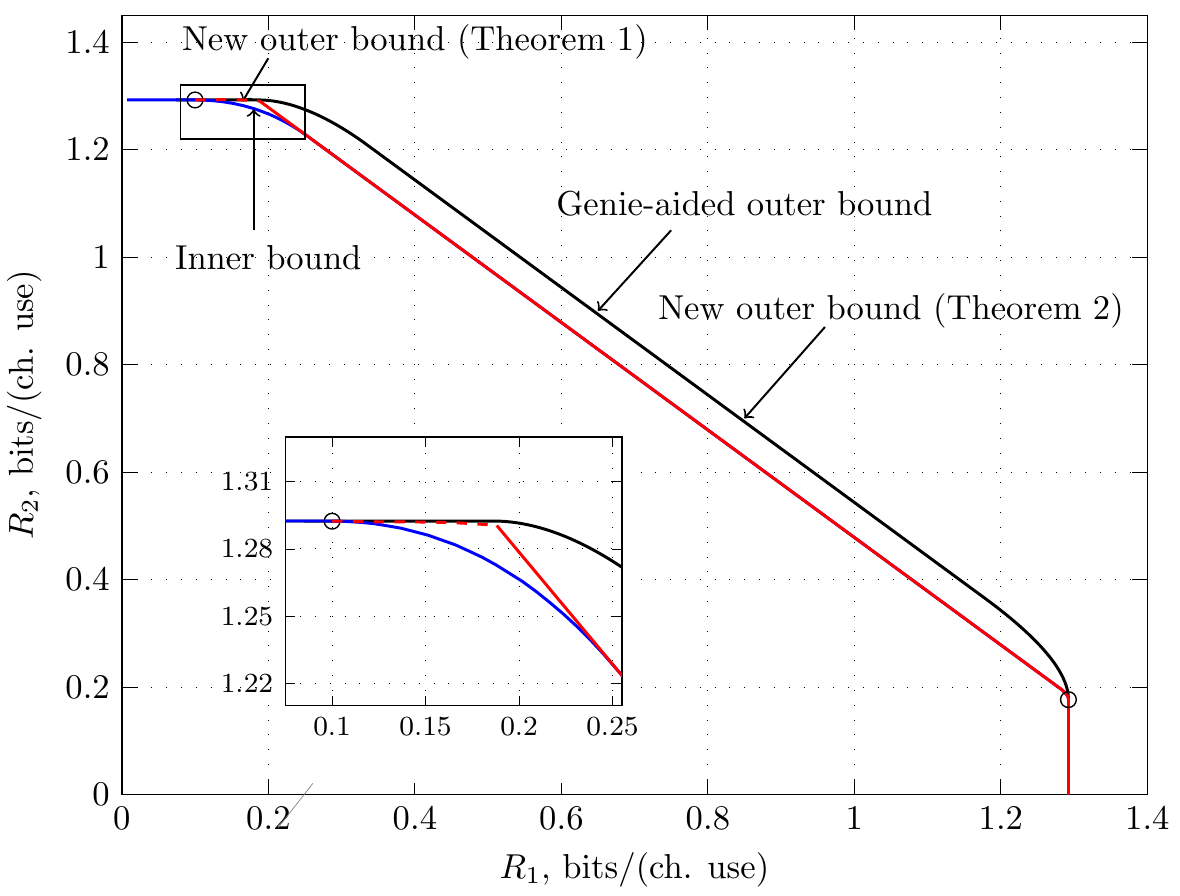}
\caption{\label{fig:ca_region_full_non} Inner and outer bounds on the capacity region region $\setC(P_1,P_2,Q)$ with $P_1=5$, $P_2=5$, and $Q=12$.}
\end{figure}

\subsubsection{Numerical results}
\label{sec:numerical_results_nondegraded}

In Fig.~\ref{fig:ca_region_full_non}, we compare our new bounds in Theorems~\ref{thm:outer-bound-mac} and~\ref{thm:outer-bound-r2} with the inner and outer bounds reviewed in Section~\ref{sec:problem-setup} for  $P_1=5$, $P_2=5$, and $Q=12$.
It is not difficult to verify that this set of parameters satisfy Condition~\ref{cond:parameters}.
We make the following observations from Fig.~\ref{fig:ca_region_full_non}.
\begin{itemize}
\item The top corner point of the capacity region is given by the rate pair $(1.29, 0.1)$. 
\item The outer bound in Theorem~\ref{thm:outer-bound-r2} matches the inner bound when \ifthenelse{\boolean{conf}}{$R_1\geq  0.25$ bits/(ch. use).}{$R_1\geq R_{1,\mathrm{th}} = 0.25$ bits/(ch. use).}
\item In the regime $R_1\in(0.1,0.25)$, there is a gap between our outer bounds and the inner bound. This regime can be further divided into two regimes: if  $R_1 \in (0.1,0.19)$, then Theorem~\ref{thm:outer-bound-mac} yields a tighter upper bound on $R_2$; if $R_1 \in (0.19,0.25)$, then the bound in Theorem~\ref{thm:outer-bound-r2}  is tighter.
\end{itemize}
%
Overall, our outer bounds provide a substantial improvement over the genie-aided outer bound in~\eqref{eq:outer-somekh1}--\eqref{eq:ob-sum-rate-mac}.

\begin{figure}
\centering
\includegraphics[scale=1]{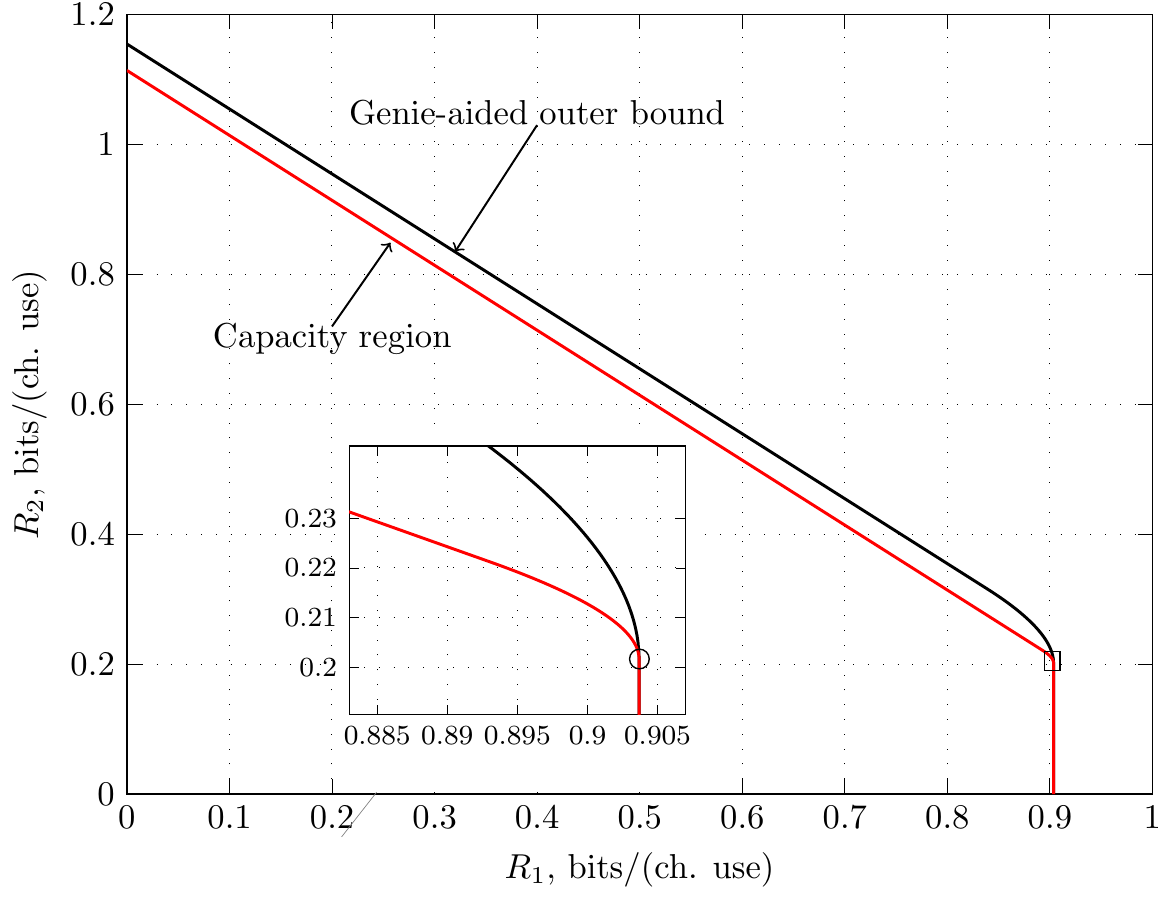}
\caption{\label{fig:ca_region_full} A comparison between the capacity region $\setC(P_1,P_2,Q)$ and the genie-aided outer bound with $P_1=2.5$, $P_2=5$, and $Q=12$.}
\end{figure}

In Fig.~\ref{fig:ca_region_full}, we consider another set of parameters with  $P_1=2.5$, $P_2=5$, and $Q=12$.
In this case,  we have $C_{\mathrm{helper}} =C_{\mathrm{sum}} =1.11$ bits/(ch. use), and the capacity region $\setC(P_1,P_2,Q)$  is completely characterized by Corollary~\ref{cor:complete-characterization-1}.
\ifthenelse{\boolean{conf}}{
We observe that the boundary of the capacity region consists of three pieces: a straight line connecting the two points $(0,C_{\mathrm{helper}})$ and $(0.89, 0.22)$, a curved line connecting $(0.89, 0.22)$ and the bottom corner point $(0.9, 0.2)$, and a vertical  line connecting the bottom corner point $(0.9, 0.2)$ and $(0.9,0)$.
}{As explained in the proof  of Corollary~\ref{cor:complete-characterization-1}, the capacity region consists of three pieces: a straight line connecting the two points $(0,C_{\mathrm{helper}})$ and $(\bar{R}_1,\bar{R}_2)$, where $\bar{R}_1=0.89$ bits/(ch. use) and $\bar{R}_2=0.22$ bits/(ch. use), a curved line connecting $(\bar{R}_1,\bar{R}_2)$ and the bottom corner point $(0.9, 0.2)$, and a vertical  line connecting the bottom corner point $(0.9, 0.2)$ and $(0.9,0)$.}
 
 \subsubsection{Generalization to MAC with non-Gaussian state}
 \label{sec:non_gaussian_states}

  In the proofs of Theorems~\ref{thm:outer-bound-mac}--\ref{thm:sum-rate-capacity}, the only place where we have used the Gaussianity of $S^n$ is to optimize appropriate mutual information terms over $P_{X_1|S}$ (see, e.g.,~\eqref{eq:intuitive-max-gaussian}).
  If the state sequence $S^n$ is non-Gaussian but is i.i.d., then the upper bound~\eqref{eq:thm1-ub} remains valid if $f(\delta)$ is replaced by
\begin{IEEEeqnarray}{rCl}
\tilde{f}(\delta) \define \max_{P_{X_1|S}} \Big\{ I(X_1+S; Y_G) - I(S; Y_{\delta}) \Big\}.\IEEEeqnarraynumspace
\end{IEEEeqnarray}
In this case, the top corner point becomes
\begin{equation} 
\widetilde{C}_1 = \max_{P_{X_1|S}} \{I(X_1+S; Y_G) - I(X_1+S; S)\}
\label{eq:corner-point-top-non-gaussian}
\end{equation}
and the  sum rate capacity becomes
\begin{IEEEeqnarray}{rCl}
C_{\mathrm{sum}} = \max_{P_{X_1|S}P_{X_2} }  \big( I(X_1;Y|X_2,S) + I(X_2;Y)\big).
\label{eq:sum-capacity-general}
\end{IEEEeqnarray} 
Furthermore, both~\eqref{eq:sum-rate-capacity} and~\eqref{eq:sum-capacity-general} can be achieved by treating interference as noise for the non-cognitive user, and by using generalized dirty paper coding for the cognitive user (recall that, in the dirty paper coding problem, the state $S$ does not need to be Gaussian; see, e.g.,~\cite[Sec.~7.7]{elgamma-11a}).

\subsection{Dirty MAC with Degraded Message Sets}

\label{sec:mac_degraded_results}


Theorem~\ref{thm:outer-bound-mac-degraded-1} below extends the outer bound in Theorem~\ref{thm:outer-bound-mac} to the dirty MAC with degraded message sets. 

\begin{thm}
\label{thm:outer-bound-mac-degraded-1}
The capacity region $\setC_{\deg}(P_1,P_2,Q)$ of the dirty MAC with degraded message sets is outer-bounded by the region with rate pairs $(R_1,R_2)$ satisfying
\begin{IEEEeqnarray}{rCl}
R_2 \leq C_{\mathrm{helper}}
\label{eq:ub-r2-trivial-degraded}
\end{IEEEeqnarray}
and
\begin{IEEEeqnarray}{rCl}
R_1\leq \min_{0\leq \delta \leq 1}  \Big\{ \frac{1}{2} \log\lefto( 1 + \frac{1+P_2-\delta}{ P_2\delta} \tilde{g}(R_2)\right) + f(\delta) \Big\} + (c_2+c_3)(C_2 - R_2) \IEEEeqnarraynumspace
\label{eq:thm1-ub-degraded}
\end{IEEEeqnarray}
where $f(\cdot)$ is defined in~\eqref{eq:def-f-delta}, 
\begin{IEEEeqnarray}{rCl}
\tilde{g}(R_2) \define \exp\lefto(2 c_2 \sqrt{C_2-R_2} + 2(C_2-R_2)\right) -1 \IEEEeqnarraynumspace
\label{eq:def-g-r2-degraded}
\end{IEEEeqnarray}
with
\begin{IEEEeqnarray}{rCl}
c_2 &\define& \frac{3\sqrt{1+(\sqrt{P_1}+\sqrt{P_2}+\sqrt{Q})^2} + 4(\sqrt{P_1} + \sqrt{Q}) }{\sqrt{(1+P_2)/(2\log e)}} \IEEEeqnarraynumspace
\label{eq:def-c2}
\end{IEEEeqnarray}
and 
\begin{IEEEeqnarray}{rCl}
c_3&\define& \sqrt{2(1+P_2)\log e} \cdot \left(3\sqrt{1+(\sqrt{P_1}+\sqrt{P_2}+\sqrt{Q})^2} + 4(\sqrt{P_1}+\sqrt{P_2} + \sqrt{Q}) \right). \IEEEeqnarraynumspace \label{eq:def-c3}
\end{IEEEeqnarray}
\end{thm}
\begin{IEEEproof}
See Section~\ref{sec:proof-corner-degraded}.
\end{IEEEproof}

As a corollary of Theorem~\ref{thm:outer-bound-mac-degraded-1}, we establish that  under Condition~\ref{cond:parameters}, the top corner point established in~\eqref{eq:corner-point-top} is unchanged even if  the non-cognitive user knows the message of the cognitive user. 
Formaly, the top corner point is defined as 
\begin{IEEEeqnarray}{rCl}
\widetilde{C}_{\mathrm{deg},1}(P_1,P_2,Q) &\define& \max \{ R_1: (R_1, C_2) \in \setC_{\deg}(P_1,P_2,Q) \} \label{eq:corner-point-r1} .\IEEEeqnarraynumspace
\end{IEEEeqnarray}

\begin{cor}
\label{thm:corner-point-result-degraded}
For the dirty MAC  with degraded message sets,  if $P_1$, $P_2$, and $Q$ satisfy Condition~\ref{cond:parameters},  then  
\begin{IEEEeqnarray}{rCl}
\widetilde{C}_{\deg,1}(P_1,P_2,Q) = f(0) 
\label{eq:corner-point-top-degraded}
\end{IEEEeqnarray}
with $f(\cdot)$ defined in~\eqref{eq:def-f-delta}.
\end{cor}

Note that, for the dirty MAC with degraded message sets, both the bottom corner point and the sum rate capacity can be established from the inner and outer bounds in~\cite{zaidi2009-06a}. 
The next theorem provides an outer bound, which is uniformly tighter than the one in~\eqref{eq:outer-zaidi} and~\eqref{eq:outer-somekh2} derived in~\cite[Th.~4]{zaidi2009-06a}.

\begin{thm}
\label{thm:mac-degraded-bound2}
The capacity region of the dirty MAC with degraded message set is outer-bounded by the region with rate pairs $(R_1,R_2)$ satisfying
\begin{IEEEeqnarray}{rcl}
R_2 &\leq&\frac{1}{2}\log(1+P_2(1-\rho_2^2)) \label{eq:out-r2-degraded-gau}\\
R_2&\leq& \frac{1}{2}\log(1+P_1(1-\rho_1^2 -\rho_s^2)) \notag\\
&&+\,\frac{1}{2}\log\mathopen\Big(1+\frac{P_2(1-\rho_2^2)  }{1+(\sqrt{Q}+\rho_s\sqrt{P_1})^2 +P_1(1-\rho_1^2 -\rho_s^2)}\Big) \label{eq:out-r2-degraded-gau2}\\
R_1+R_2&\leq& \frac{1}{2}\log(1+P_1(1-\rho_1^2 -\rho_s^2)) \notag\\
&&+\,\frac{1}{2}\log\mathopen\Big(1+\frac{P_2(1-\rho_2^2) +(\rho_2\sqrt{P_2} +\rho_1\sqrt{P_1})^2}{1+(\sqrt{Q}+\rho_s\sqrt{P_1})^2 +P_1(1-\rho_1^2 -\rho_s^2)}\Big)\IEEEeqnarraynumspace  \label{eq:out-sum-degraded-gau}
\end{IEEEeqnarray}
for some $\rho_1\in[0,1]$, $\rho_2\in[0,1]$, $\rho_s \in [-1,0]$ that satisfy
\begin{IEEEeqnarray}{rCl}
\rho_1^2+\rho_s^2 \leq 1.
\label{eq:cond-rho1-rhos}
\end{IEEEeqnarray}
\end{thm}

\begin{IEEEproof}
The proof of Theorem~\ref{thm:mac-degraded-bound2} follows from the following single-letter outer bound on the capacity region, whose proof is given in Section~\ref{prop-outer-bound-degraded-mac-sum}.
\begin{prop}
\label{prop:outer-bound-degraded}
The capacity region of the dirty MAC with degraded message set is outer-bounded by the region with rate pairs $(R_1,R_2)$ satisfying
\begin{IEEEeqnarray}{rCl}
R_2 &\leq& I(X_2;Y|X_1,S,U) \label{eq:out-prop-degraded-r2}\\
R_2&\leq& I(X_1;Y|X_2,S,U) + I(X_2;Y|U)  \label{eq:out-prop-degraded-r22}\\
R_1+R_2 &\leq& I(X_1;Y|X_2,S,U) + I(X_2,U;Y) \IEEEeqnarraynumspace \label{eq:out-sum-degraded}
\end{IEEEeqnarray}
for some joint distributions $P_{X_1,X_2,S,U}$ that satisfy 
\begin{itemize}
\item $X_1$ and $X_2$ are conditionally independent given $U$;
\item $U$ and $X_2$ are independent of $S$;
\item $\Ex{}{X_1^2} \leq P_1$ and $\Ex{}{X_2^2}\leq P_2$.  
\end{itemize}
\end{prop}

To prove Theorem~\ref{thm:mac-degraded-bound2}, it remains to show that the bounds in~\eqref{eq:out-prop-degraded-r2}--\eqref{eq:out-sum-degraded} are  maximized when $U$, $S$, $X_1$, and $X_2$ are jointly Gaussian. 
The proof of this result is provided in the appendix.
\end{IEEEproof}
 
Next, we explain how the outer bound in Proposition~\ref{prop:outer-bound-degraded} improves upon~\eqref{eq:zaidi-09-rate2} and~\eqref{eq:zaidi-09-sum-rate}. 
Observe that~\eqref{eq:zaidi-09-sum-rate} can be rewritten as
\begin{IEEEeqnarray}{rCl}
R_1+R_2\leq I(X_1;Y|S,X_2) + I(X_2;Y) 
\end{IEEEeqnarray}
where the joint probability distribution of  $S$, $X_1$, and $X_2$ has the form $P_{S}P_{X_2} P_{X_1|X_2,S}$. 
The key difference between Proposition~\ref{prop:outer-bound-degraded} and the outer bound in~\eqref{eq:zaidi-09-rate2} and~\eqref{eq:zaidi-09-sum-rate} is the introduction of the auxiliary random variable $U$ in Proposition~\ref{prop:outer-bound-degraded}.
The intuition for this auxiliary random variable is as follows. 
Since the non-cognitive user knows both messages $\msg_1$ and $\msg_2$, its  input $X_2$ must contain two parts, where each part depends only on one message. 
The auxiliary random variable $U$ in Proposition~\ref{prop:outer-bound-degraded} captures precisely the part of $X_2$ that depends on $M_1$. 
Since the input $X_1$ of the cognitive user depends on $X_2$ only through the message $M_1$, and hence through $U$, 
we see that  $X_1$ and $X_2$ are conditionally independent given $U$, as stated in the proposition. 
For comparison, the bounds~\eqref{eq:zaidi-09-rate2} and~\eqref{eq:zaidi-09-sum-rate}, which allow arbitrary dependence between $X_1$ and $X_2$, is looser than the bound in Proposition~\ref{prop:outer-bound-degraded} (unless $R_2=0$, in which case $U=X_2$).


\begin{figure}
\centering
\includegraphics[scale=1]{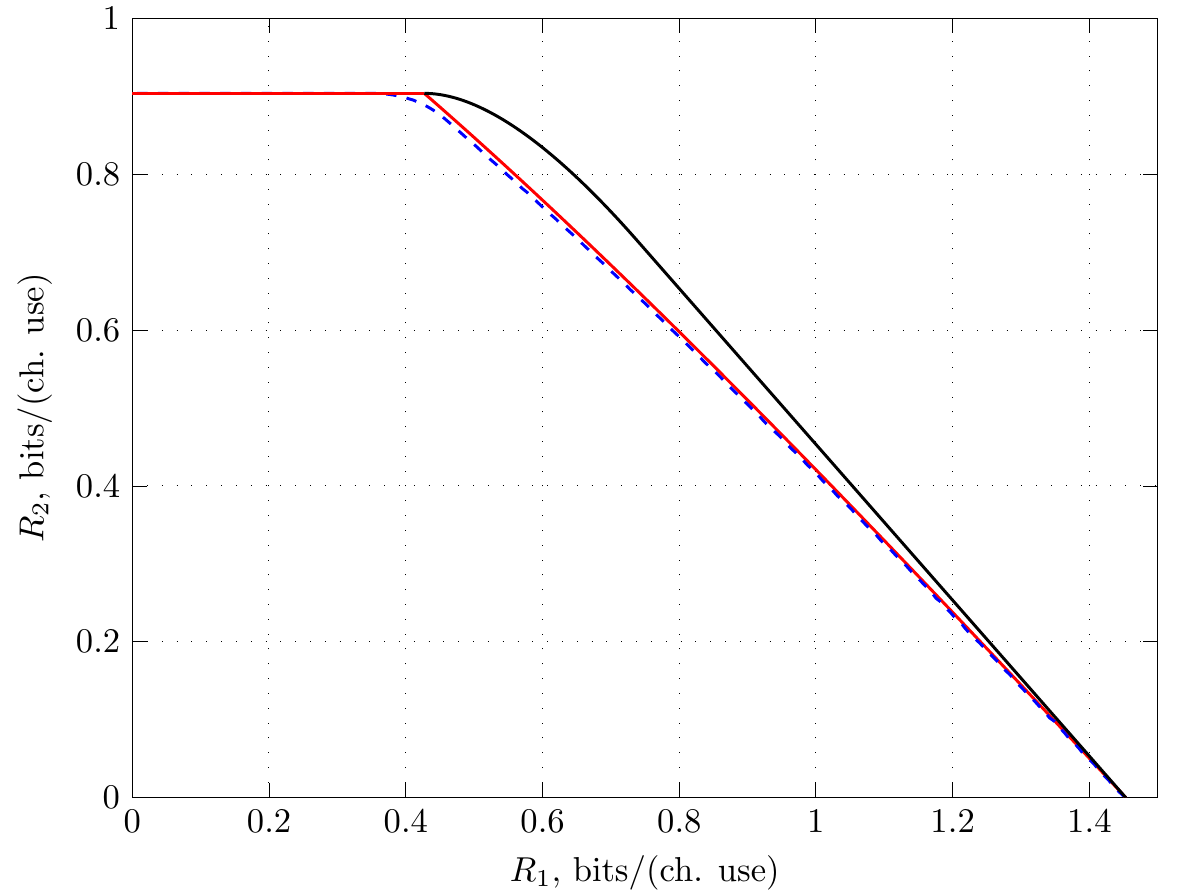}
\caption{Inner and outer bounds for the capacity region of the dirty MAC with degraded message sets for $P_1=4$, $P_2=2.5$, and $Q=5$. The red solid curve denotes our new outer bound in Theorem~\ref{thm:mac-degraded-bound2}, the blue dashed curve and the black curve denote the inner and outer bounds obtained in~\cite{zaidi2009-06a}.
\label{fig:region-outer-degraded1}}
\end{figure}
\begin{figure}
\centering
\includegraphics[scale=1]{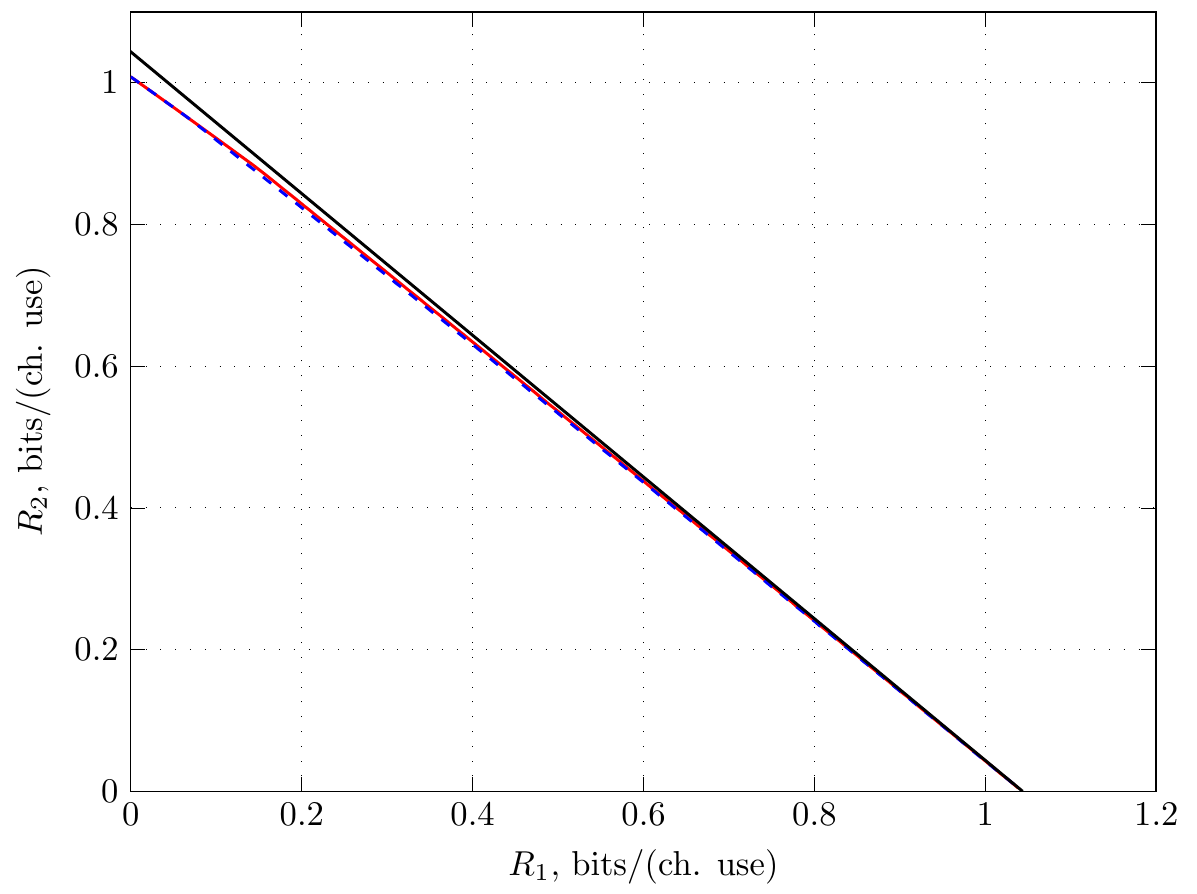}
\caption{Inner and outer bounds for the capacity region of the dirty MAC with degraded message sets for $P_1=2$, $P_2=5$, and $Q=12$. The red solid curve denotes our new outer bound in Theorem~\ref{thm:mac-degraded-bound2}, the blue dashed curve and the black curve denote the inner and outer bounds obtained in~\cite{zaidi2009-06a}.
\label{fig:region-outer-degraded2}
}
\end{figure}

In Figs.~\ref{fig:region-outer-degraded1} and~\ref{fig:region-outer-degraded2}, we  compare our new outer bound in Theorem~\ref{thm:mac-degraded-bound2} with the inner and outer bounds in~\cite{zaidi2009-06a} for different values of $P_1$, $P_2$, and $Q$.
In both figures, the red solid curve denotes our new outer bound in Theorem~\ref{thm:mac-degraded-bound2}, and the blue dashed curve and the black curve denote the inner and outer bounds obtained in~\cite{zaidi2009-06a}.
As expected, our new outer bound is tighter than the outer bound in~\cite[Th.~4]{zaidi2009-06a}, and is almost on top of the inner bound for the parameters considered in Figs.~\ref{fig:region-outer-degraded1} and~\ref{fig:region-outer-degraded2}.
For the scenario considered in Fig.~\ref{fig:region-outer-degraded1},  our outer bound does not match the inner bound (unless $R_2 =0$).  Numerically, we observe that the gap between the inner bound and our outer bound is less than $0.013$ bits$/$(ch. use).
For the scenario considered in Fig.~\ref{fig:region-outer-degraded2},  our outer bound matches the inner bound if either $R_1\leq 0.1$ or $R_2=0$. 
The gap between the inner and outer bounds in this scenario is less than $3.4\times 10^{-3}$ bits/(ch. use). 
%


%
%
%

\subsection{The helper problem}
\label{sec:helper-problem-results}
The outer bound in Theorem~\ref{thm:outer-bound-mac} also yields an upper bound on the capacity of the helper problem as shown in the next result. 

\begin{thm}
\label{thm:helper}
For the helper problem, we have 
\ifthenelse{\boolean{singcol}}{
\begin{IEEEeqnarray}{rCl}
C_{\mathrm{helper}} \leq  \max\mathopen{}\bigg\{R_2: R_2 \leq C_2, \text{ and } \min_{0\leq \delta \leq 1}  \Big\{ \frac{1}{2} \log\lefto( 1 + \frac{1+P_2-\delta}{ P_2\delta} g(R_2)\right) + f(\delta) \Big\}  \geq 0 \bigg\} \IEEEeqnarraynumspace
\label{eq:helper-upper-bound}
\end{IEEEeqnarray}
}{
\begin{IEEEeqnarray}{rCl}
&&C_{\mathrm{helper}} \leq  \max\mathopen{}\bigg\{R_2: R_2 \leq C_2, \text{ and }\notag\\ 
&& \min_{0\leq \delta \leq 1}  \Big\{ \frac{1}{2} \log\lefto( 1 + \frac{1+P_2-\delta}{ P_2\delta} g(R_2)\right) + f(\delta) \Big\}  \geq 0 \bigg\} \IEEEeqnarraynumspace
\label{eq:helper-upper-bound}
\end{IEEEeqnarray}}
where $g(\cdot)$ and $f(\cdot)$ are defined in~\eqref{eq:def-g-r2} and~\eqref{eq:def-f-delta}, respectively.
\end{thm}

\begin{IEEEproof}
Setting $R_1=0$ in the outer bound~\eqref{eq:thm1-ub} in Theorem~\ref{thm:outer-bound-mac}, we conclude that the rate $R_2$ of the non-cognitive user must satisfy
\begin{IEEEeqnarray}{rCl}
 \min_{0\leq \delta \leq 1}  \Big\{ \frac{1}{2} \log\lefto( 1 + \frac{1+P_2-\delta}{ P_2\delta} g(R_2)\right) + f(\delta) \Big\}  \geq 0. \IEEEeqnarraynumspace
\end{IEEEeqnarray}
This implies~\eqref{eq:helper-upper-bound}.
\end{IEEEproof}

A simple consequence of Theorem~\ref{thm:helper} is the following result, which shows that Condition~\ref{cond:parameters} is both necessary and sufficient  for the non-cognitive user to achieve the channel capacity without state dependence.
\begin{cor}
\label{cor:helper-equivalent}
For the helper problem, the following two statements are equivalent:
\begin{enumerate}
\item $C_{\mathrm{helper}} = \frac{1}{2}\log (1+P_2)$;
\item The channel parameters $P_1$, $P_2$, and $Q$ satisfy Condition~\ref{cond:parameters};
\item $f(0) \geq 0$, where $f(\cdot)$ is defined in~\eqref{eq:def-f-delta}.
\end{enumerate}
\end{cor}

\begin{figure}
 \centering
\includegraphics[scale=1]{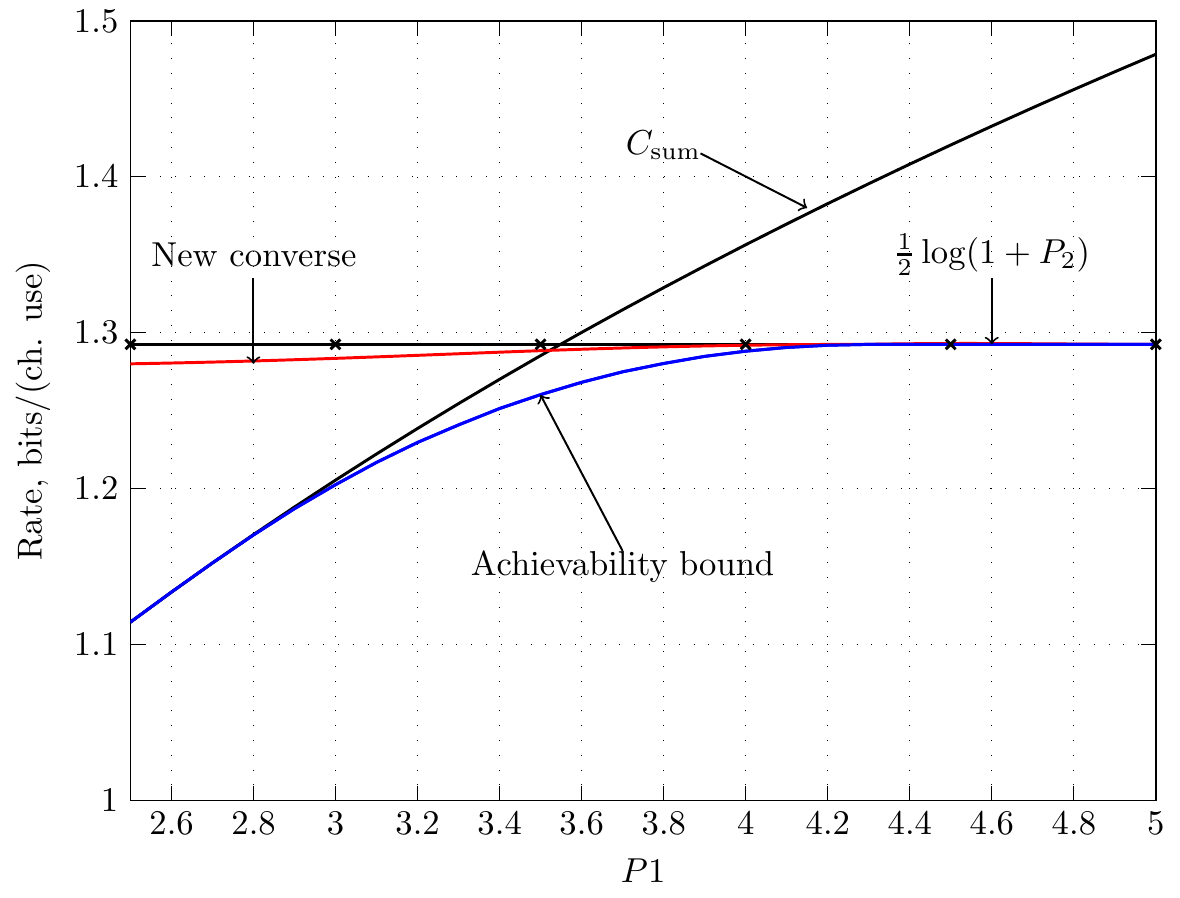}
\caption{Upper and lower bounds on $C_\mathrm{helper}$ as a function of $P_1$ for $P_2=5$ and $Q=12$.\label{fig:helper-capacity}}
\end{figure}

In Fig.~\ref{fig:helper-capacity}, we compare the new upper bound in Theorem~\ref{thm:helper} with the upper and lower  bounds in~\cite{sun2016-12a}. 
The two upper bounds reported in~\cite[Lemmas 2 and 3]{sun2016-12a} correspond to
\begin{IEEEeqnarray}{rCl}
C_{\mathrm{helper}} \leq C_{\mathrm{sum}}
\label{eq:ub-helper-csum}
\end{IEEEeqnarray}
and
\begin{IEEEeqnarray}{rCl}
C_{\mathrm{helper}} \leq \frac{1}{2}\log(1+P_2)
\label{eq:ub-helper-c2}
\end{IEEEeqnarray}
respectively. 
The lower bound (achievability bound) is~\cite[Th.~1]{sun2016-12a}.
As observed in~\cite{sun2016-12a},  the upper bound~\eqref{eq:ub-helper-csum} is tight (i.e., $C_{\mathrm{helper}} = C_{\mathrm{sum}}$) if $P_1 \leq 2.5$, and 
the bound~\eqref{eq:ub-helper-c2} is tight  (i.e., $C_{\mathrm{helper}} =  \frac{1}{2}\log(1+P_2)$) if $P_1\geq 4.5$. 
Our new upper bound is tighter than~\eqref{eq:ub-helper-csum} and~\eqref{eq:ub-helper-c2} for $P_1\in [3.5,4.5]$.

\section{Technical Proofs}

 \subsection{Proof of Theorem~\ref{thm:outer-bound-mac}}
\label{sec:proof-thm-polywu}
The upper bound~\eqref{eq:ub-r2-trivial} is straightforward. 
The proof of~\eqref{eq:thm1-ub}, which builds upon the intuition described in Section~\ref{sec:outer-bounds-mac},  consists of four steps.
\begin{enumerate}
\item \label{item:setp1}We derive an upper bound on 
\begin{equation}
 I_\delta \define I(X_1^n+S^n; Y_\delta^n) - I(X_1^n+S^n; Y^n_G) 
\label{eq:def-g-difference-two-mutual-info}
\end{equation}
that holds for all $X_1^n(\msg_1,S^n)$ such that the uninformed user is able to communicate at rate $R_2$ with vanishing  error probability. 
Here, $Y_G^n$ and $Y_\delta^n$ are defined in~\eqref{eq:def-yg-n} and~\eqref{eq:def-ydelta-n}, respectively.
The derivation relies on an elegant argument of Polyanskiy and Wu~\cite{polyanskiy16-07a}, used in the derivation of  the outer bound on the capacity region of Gaussian interference channels.  
\item\label{item:setp2} We obtain a lower bound on $I_\delta$ that involves $R_1$.  
Combining this lower bound with the upper bound obtained in the first step, we obtain a multi-letter upper bound on $R_1$ that depends on the joint distribution of $X_1^n$ and $S^n$ but not on $X_2^n$.

\item\label{item:setp3} We single-letterize the upper bound obtained in Step~\ref{item:setp2}.

\item\label{item:setp4} We show that the upper bound obtained in Step~\ref{item:setp3} is maximized when $X_1$ and $S$ are jointly Gaussian. 

\end{enumerate}

\subsubsection{Step 1: Upper-bounding $I_\delta$}
\label{sec:upper_bound_idelta}
The derivation follows closely the proof of~\cite[Th.~7]{polyanskiy16-07a}.
Let 
 \begin{IEEEeqnarray}{rCl}
R_1 &\define& \frac{1}{n}I(\msg_1;Y^n) \label{eq:def-rate-fano-1}
\\
\label{eq:def-rate-fano}
R_2 &\define& \frac{1}{n} I(X_2^n;Y^n).
\end{IEEEeqnarray}
As explained in~\cite{polyanskiy16-07a},  this definition of rate agrees with the operational definition (i.e., the ratio between the logarithm of the number of messages and the blocklength) asymptotically.  
Without loss of generality, we assume that $X_1^n$ and $X_2^n$ have zero mean.  
Let 
\begin{IEEEeqnarray}{rCl}
N_{S}(\gamma) \define \exp\lefto\{\frac{2}{n} h(X_1^n +S^n +\sqrt{\gamma} Z^n)\right\}
\end{IEEEeqnarray}
where $Z^n \sim \realg(0,\matI_n)$ is independent of $X_1^n$ and $S^n$.
By Costa's entropy power inequality~\cite{costa1985-11a},  the function $  N_{S}(\cdot)$ is concave. 
The term $I_\delta$ in~\eqref{eq:def-g-difference-two-mutual-info} can be expressed in terms of $N_S(\cdot)$ as
\begin{IEEEeqnarray}{rCl}
I_\delta = \frac{n}{2}\log\frac{N_S(\delta)}{N_S(1+P_2)} + \frac{n}{2}\log \frac{1+P_2}{\delta}.
\label{eq:def-I-delta-n}
\end{IEEEeqnarray}
Repeating the steps in~\cite[Eqs.~(41)--(43)]{polyanskiy16-07a}, we obtain (recall that $G^n\sim \realg(0, P_2\matI_n)$)
\begin{IEEEeqnarray}{rCl}
D(P_{X_2^n +Z^n} \| P_{G^n +Z^n}) \leq n (C_2 - R_2)
\label{eq:bound-KL-x2-G}
\end{IEEEeqnarray}
where $D(\cdot\|\cdot)$ denotes the relative entropy between two distributions, and
\begin{IEEEeqnarray}{rCl}
nR_2 &=& I(X_2^n;Y^n)\\
&=& h(Y^n)- h(Y^n_G)  + h(Y_G^n)- h(X_1^n + S^n + Z^n) \IEEEeqnarraynumspace\\
&=&h(Y^n)- h(Y^n_G)   + \frac{n}{2}\log\frac{N_S(1+P_2)}{N_S(1)}. 
\label{eq:express-r2-four-entrory}
\end{IEEEeqnarray}
Note that $\Ex{}{X_1^n +S^n} = \mathbf{0}$, $\Ex{}{X_2^n} = \mathbf{0}$, $\Ex{}{\|X_2^n\|^2} \leq nP_2$, and 
\begin{IEEEeqnarray}{rCl}
\IEEEeqnarraymulticol{3}{l}{
\Ex{}{\|X_1^n +S^n\|^2} }\notag\\
\quad &=& \Ex{}{\|X_1^n\|^2} + \Ex{}{\|S^n\|^2} + 2 \Ex{}{\langle X_1^n, S^n\rangle}  \IEEEeqnarraynumspace\\
&\leq& n P_1 + nQ +  2\Ex{}{\|X_1^n\| \|S^n\|} \\
&\leq& nP_1 + nQ + 2\sqrt{\Ex{}{ \|X_1^n\|^2} \Ex{}{\|S^n\|^2}} \\
&\leq& n(\sqrt{P_1} +\sqrt{Q} )^2.
\label{eq:bound-x1n-norm}
\end{IEEEeqnarray}
By~\cite[Prop. 2]{polyanskiy16-07a}, the random variable $Y_G^n$ is $(\frac{3\log e}{1+P_2} , \frac{4(\sqrt{P_1} +\sqrt{Q}) \log e }{1+P_2})$-regular, i.e., the probability density function $p_{Y_G^n}(y^n)$ of $Y_G^n$  satisfies 
\ifthenelse{\boolean{singcol}}{
\begin{IEEEeqnarray}{rCl}
\|\gradient \log p_{Y_G^n} (y^n)\| &\leq& \frac{3\log e}{1+P_2}  \|y^n\| + \frac{4(\sqrt{P_1} +\sqrt{Q}) \log e }{1+P_2},  \quad \forall y^n\in\mathbb{R}^n.
\end{IEEEeqnarray}
}{
\begin{IEEEeqnarray}{rCl}
\|\gradient \log p_{Y_G^n} (y^n)\| &\leq& \frac{3\log e}{1+P_2}  \|y^n\| + \frac{4(\sqrt{P_1} +\sqrt{Q}) \log e }{1+P_2}, \notag\IEEEeqnarraynumspace\\
&&\qquad\qquad\qquad \qquad\qquad\forall y^n\in\mathbb{R}^n.
\end{IEEEeqnarray}}
Therefore, by~\cite[Prop. 1]{polyanskiy16-07a}, the entropy difference between $Y^n$ and $Y_G^n$ can be bounded via the Wasserstein distance $W_2(P_{Y^n} , P_{Y_G^n})$ (see~\cite[p.~12]{villani03-a} for the definition of $W_2$) as 
\begin{IEEEeqnarray}{rCl}
\IEEEeqnarraymulticol{3}{l}{
h(Y^n)- h(Y^n_G) }\notag\\
\quad & \leq  &\left( 3\sqrt{1+(\sqrt{P_1}+\sqrt{Q})^2+P_2} + 4(\sqrt{P_1} + \sqrt{Q}) \right) \notag\\
&& \cdot \, \frac{ \sqrt{n} \log e}{1+P_2} \cdot W_2(P_{Y^n} \| P_{Y_G^n}). 
\label{eq:diff-entropy-yn-ygn}
\end{IEEEeqnarray} 
Furthermore, we have 
\begin{IEEEeqnarray}{rCl}
W_2(P_{Y^n} \| P_{Y_G^n})&\leq& W_2(P_{X_2^n +Z^n} \| P_{G^n +Z^n})\label{eq:bound-wasser-1} \\
&\leq&\sqrt{\frac{2(1+P_2)}{\log e} D(P_{X^n_2 +Z^n} \| P_{G^n+Z^n} )} \label{eq:bound-wasser-2} \IEEEeqnarraynumspace\\
&\leq& \sqrt{\frac{2n (1+P_2)}{\log e}  (C_2 - R_2)}. \label{eq:bound-wasser-3}
\end{IEEEeqnarray}
Here,~\eqref{eq:bound-wasser-1} follows because the $W_2(\cdot,\cdot)$ distance is non-decreasing under convolutions,~\eqref{eq:bound-wasser-2} follows by using Talagrand's inequality~\cite{talagrand1996-05a},  and~\eqref{eq:bound-wasser-3} follows from~\eqref{eq:bound-KL-x2-G}. 
Substituting~\eqref{eq:bound-wasser-3} into~\eqref{eq:diff-entropy-yn-ygn}, and then~\eqref{eq:diff-entropy-yn-ygn} into~\eqref{eq:express-r2-four-entrory}, we conclude that
\begin{equation} 
\log\frac{N_S(1)}{N_S(1+P_2)} \leq 2 c_1\sqrt{ C_2-R_2 }  +  2(C_2 - R_2)
-\log(1+P_2)
\label{eq:bound-on-ratio-ns-initial}
\end{equation}
where $c_1$ is defined in~\eqref{eq:def-c1}, or equivalently,
\begin{IEEEeqnarray}{rCl}
\frac{N_S(1)}{N_S(1+P_2)}  \leq \frac{\exp\lefto(2 c_1\sqrt{C_2-R_2} + 2(C_2-R_2) \right) }{1+P_2}. \IEEEeqnarraynumspace
\end{IEEEeqnarray}
Let $\alpha \define P_2/(1+P_2 -\delta)$ be such that  
\begin{IEEEeqnarray}{rCl}
\alpha \delta + (1-\alpha)(1+P_2) = 1.
\end{IEEEeqnarray}
By the concavity of $N_S(\cdot)$, we have
\begin{IEEEeqnarray}{rCl}
\alpha N_S(\delta)  + (1-\alpha) N_S(1+P_2) &\leq& N_S(1)
\end{IEEEeqnarray}
which implies that
\begin{IEEEeqnarray}{rCl}
\IEEEeqnarraymulticol{3}{l}{
\frac{N_S(\delta)}{N_S(1+P_2)}  }\notag\\
&\leq& \frac{1}{\alpha}\frac{ N_S(1) - (1-\alpha) N_S(1+P_2) }{N_S(1+P_2)}\\
&\leq&\frac{1}{\alpha} \left( \frac{\exp\lefto(2 c_1\sqrt{C_2-R_2} + 2(C_2-R_2) \right) }{1+P_2} - 1 +\alpha \right). \IEEEeqnarraynumspace
\label{eq:ratio-N_S-bound}
\end{IEEEeqnarray}
Substituting~\eqref{eq:ratio-N_S-bound} into~\eqref{eq:def-I-delta-n}, we conclude that 
\begin{IEEEeqnarray}{rCl}
I_\delta &\leq& \frac{n}{2} \log\lefto( 1 + \frac{1+P_2-\delta}{ P_2\delta} g(R_2) \right) 
\label{eq:upper-bound-I-delta-final}
\end{IEEEeqnarray}
where $g(R_2)$ is defined in~\eqref{eq:def-g-r2}. 

\subsubsection{Step 2:  Lower-bounding $I_\delta$}
\label{sec:step2-lower-bound-idelta}
We next derive a lower bound on $I_\delta$.
Consider the following chain of (in)equalities:
\begin{IEEEeqnarray}{rCl}
I_\delta &=& I(X_1^n +S^n; Y_\delta^n) - I(X_1^n+S^n;Y^n_G) \\
&=& I(X_1^n, S^n; Y_\delta^n)  - I(X_1^n+S^n;Y^n_G) \label{eq:lower-bound2}\\
&=& I(X_1^n, S^n; Y_\delta^n, \msg_1) - I(X_1^n, S^n;\msg_1 | Y_\delta^n)\notag\\
&& -\, I(X_1^n+S^n;Y^n_G) \label{eq:lower-bound3}\\
&=& I(X_1^n, S^n; \msg_1) + I(X_1^n, S^n; Y_\delta^n| \msg_1) \notag\\
&&-\, H(\msg_1|Y_\delta^n) - I(X_1^n+S^n;Y^n_G) \label{eq:lower-bound4}\\
&=& nR_1 + I(S^n; Y_\delta^n| \msg_1) +  I(X^n_1; Y_\delta^n| S^n, \msg_1) \notag\\
&& -\, H(\msg_1|Y_\delta^n) - I(X_1^n+S^n;Y^n_G) \label{eq:lower-bound5}\\
&=& nR_1  + I(S^n; Y_\delta^n, \msg_1)   - H(\msg_1|Y_\delta^n) \notag\\
&&-\, I(X_1^n+S^n;Y^n_G) \label{eq:lower-bound6}\\
&\geq&  nR_1 +  I(S^n; Y_\delta^n) - H(\msg_1|Y_\delta^n) \notag\\
&& -\, I(X_1^n+S^n;Y^n_G). \label{eq:lower-bound7}
\end{IEEEeqnarray}
Here,~\eqref{eq:lower-bound2} follows because $(X_1^n,S^n) \to X_1^n+S^n \to Y^n_\delta$ forms a Markov chain;~\eqref{eq:lower-bound4} follows because $H(\msg_1|X^n_1,S^n,Y_\delta^n)=0$; and finally,~\eqref{eq:lower-bound6} follows because~$S^n$ is independent of $\msg_1$. 

Observe now that the channel $M_1 \to Y^n$  is stochastically degraded with respect to the channel $M_1\to Y_{\delta}^n$,  since $Y^n$ has the same distribution as 
$Y_\delta^n + X_2^n + \sqrt{1-\delta^2} \tilde{Z^n}$, where $\tilde{Z}^n\sim\realg(0,\matI_n)$.
This implies that a receiver that observes $Y_\delta^n$ is able to decode $M_1$ with vanishing error probability. By Fano's inequality, 
\begin{IEEEeqnarray}{rCl}
H(\msg_1|Y^n_\delta) = o(n). 
\label{eq:fano-on-delta-channel}
\end{IEEEeqnarray}
Here, the $\littleo(n)$ term depends on $R_1$ and the error probability of the cognitive encoder, but not on the joint probability distribution of $X_1^n$ and $S^n$. 
Using~\eqref{eq:fano-on-delta-channel} in~\eqref{eq:lower-bound7} we obtain that
\begin{IEEEeqnarray}{rCl}
I_\delta  \geq nR_1 + I(S^n; Y_\delta^n) - I(X_1^n+S^n;Y^n_G)  + \littleo(n). \IEEEeqnarraynumspace
\label{eq:final-lower-bound-I-delta}
\end{IEEEeqnarray}
Combining the lower bound~\eqref{eq:final-lower-bound-I-delta} with the upper bound~\eqref{eq:upper-bound-I-delta-final}, we conclude that
\begin{IEEEeqnarray}{rCl}
nR_1 &\leq & I(X_1^n+S^n;Y^n_G) - I(S^n; Y_\delta^n)  \notag\\ 
&&+ \, \frac{n}{2} \log\lefto( 1 + \frac{1+P_2-\delta}{ P_2\delta} g(R_2)\right)    +  \littleo(n).
\label{eq:ub-r1-max-original}
\end{IEEEeqnarray}
It remains to upper-bound the first two terms on the  RHS of~\eqref{eq:ub-r1-max-original}. This is done in the next two sections.

\subsubsection{Step 3: Single-letterization}
\label{sec:single-letterization-proof-thm1}

%
Observe that  
\begin{IEEEeqnarray}{rCl}
\IEEEeqnarraymulticol{3}{l}{
I(X_1^n+S^n; Y^n_G) }\notag\\
\quad &=&  \sum\limits_{i=1}^{n} \left( h(Y_{G,i} | Y_G^{i-1}) - h(Y_{G,i} | X_{1,i},S_i) \right)\\
&\leq& \sum\limits_{i=1}^{n} \left( h(Y_{G,i}) - h(Y_{G,i} | X_{1,i},S_i)\right)\label{eq:up-mi-memoryless-in}\\
&=& \sum\limits_{i=1}^{n} I(X_{1,i}+S_i ; Y_{G,i})\label{eq:up-mi-memoryless}
\end{IEEEeqnarray}
and
\begin{IEEEeqnarray}{rCl}
I(S^n; Y_\delta^n) &=& h(S^n) - h(S^n|Y_{\delta}^n)\label{eq:lb-mi-memoryless-begin} \\
&=& \sum\limits_{i=1}^{n} \left( h(S_i) - h(S_i| Y_{\delta}^n, S^{i-1}) \right)\\
&\geq& \sum\limits_{i=1}^{n} \left( h(S_i) - h(S_i| Y_{\delta,i}) \right)\label{eq:lb-mi-memoryless-in}\\
&=& \sum\limits_{i=1}^{n} I(S_i;Y_{\delta, i})\label{eq:lb-mi-memoryless}
\end{IEEEeqnarray}
where both~\eqref{eq:up-mi-memoryless-in} and~\eqref{eq:lb-mi-memoryless-in} follow because conditioning reduces entropy. 
Combining~\eqref{eq:up-mi-memoryless} and~\eqref{eq:lb-mi-memoryless}, we obtain
\begin{IEEEeqnarray}{rCl}
\IEEEeqnarraymulticol{3}{l}{
 I(X_1^n+S^n; Y^n_G) -  I(S^n; Y_\delta^n)  }\notag\\
 \qquad &\leq& 
  \sum\limits_{i=1}^{n}\left( I(X_{1,i}+S_i ; Y_{G,i} ) -  I(S_i ; Y_{\delta,i})\right)  \label{eq:opt-diff-mi}
\end{IEEEeqnarray}
where the RHS of~\eqref{eq:opt-diff-mi} depends on $P_{X^n_1|S^n}$ only through the (marginal) conditional distributions $\{P_{X_{1,i}|S_i}\}$.

Now, a critical observation is that the functional $P_{X_1|S}\mapsto I(X_{1}+S ; Y_{G} ) -  I(S ; Y_{\delta})  $ is concave (recall that $Y_G$ and $Y_\delta$ are defined in~\eqref{eq:def-yg-sing} and~\eqref{eq:def-ydelta-sing}, respectively). 
This follows because, for a fixed channel, mutual information is concave in the input distribution, and for a fixed input distribution, mutual information is convex in the channel (see, e.g.,~\cite[Th. 2.7.3]{cover06-a}).  
Furthermore, both the state sequence $S^n$ and noise sequence $Z^n$ are i.i.d.. 
This allows us to conclude that 
\begin{IEEEeqnarray}{rCl}
\IEEEeqnarraymulticol{3}{l}{
  I(X_1^n+S^n; Y^n_G) -  I(S^n; Y_\delta^n)    }\notag\\
&&\leq n \max_{P_{X_1|S}: \Ex{}{X_1^2} \leq P_1}\Big\{ I(X_1+S;Y_G) - I(S; Y_\delta)\Big\}.
\label{eq:max-single-letter}
\end{IEEEeqnarray}
%
%
%

\subsubsection{Optimality of Gaussian inputs}
\label{sec:gaussian-optimal-proof-th1}
As explained in the intuitive argument after Theorem~\ref{thm:outer-bound-mac}, we will invoke the Gaussian saddle-point property to solve the maximization problem in~\eqref{eq:max-single-letter}. Lemma~\ref{lemma:gaussian-worst} below generalizes the well-known worst-case Gaussian noise result~\cite{ihara1978-04a,diggavi2001-11a} to the case in which the noise and the Gaussian input are dependent.

%

\begin{lemma}[{\!\!\cite[Th.~1]{hassibi03}}]
\label{lemma:gaussian-worst}
Let $\randvecx_{G} \sim \realg (\mathbf{0},  \matK_{x})$ and $\randvecz_{G} \sim \realg(\mathbf{0}, \matK_z)$ be Gaussian random vectors in $\realset^d$. Let $\randvecz$ be a random vector in $\realset^d$ with the same covariance matrix as $\randvecz_G$. 
Assume 
that $\randvecx_G$ is independent of $\randvecz_G$, and that 
\begin{IEEEeqnarray}{rCl}
\Ex{}{\randvecx_G\randvecz^{\mathrm{T}}} =  \mathbf{0}_{d\times d}
\label{eq:lemma-covariance-condition}
\end{IEEEeqnarray}
where the superscript $(\cdot)^{\mathrm{T}}$ denotes transposition. 
Then
\begin{equation}
I(\randvecx_G ; \randvecx_G + \randvecz_G) \leq I(\randvecx_G ; \randvecx_G + \randvecz).
\end{equation} 
\end{lemma}


We proceed as follows. For a given $P_{X_1|S}$,  let  $\rho \define \Ex{}{X_1 S}/\sqrt{P_1Q}$ be the correlation coefficient between $X_1$ and $S$. 
Denote  
\begin{IEEEeqnarray}{rCl}
\widetilde{X}_{1}  &\define& X_1- \rho\sqrt{P_1/Q} S  \\
\widetilde{S} &\define& (1 + \rho\sqrt{P_1/Q}) S.
\end{IEEEeqnarray}
It is not difficult to verify that $\Ex{}{\widetilde{X}_{1}  \widetilde{S}} =0$ and 
$\widetilde{X}_{1} + \widetilde{S} = X_1 +S$.
Therefore, we have
\begin{IEEEeqnarray}{rCl}
I(X_1+S; Y_G) &=& I(\widetilde{X}_{1} + \widetilde{S}; \widetilde{X}_{1} + \widetilde{S} + \sqrt{1+P_2} Z)  \IEEEeqnarraynumspace
\label{eq:equiv-widt-mi}
\end{IEEEeqnarray}
and
\begin{IEEEeqnarray}{rCl}
I(S; Y_\delta)  &\geq&  I(\widetilde{S}; Y_\delta) =  I(\widetilde{S}; \widetilde{S} +  \widetilde{X}_{1} +  \sqrt{\delta} Z)\IEEEeqnarraynumspace
\label{eq:lower-bound-isn-ydelta}
\end{IEEEeqnarray}
where the inequality holds with equality if $\rho\sqrt{P_1/Q} \neq -1$.  

Observe now that, for a fixed $\rho$ and $b \define \randmatE\mathopen{}\big[\widetilde{X}^2_1\big]$, the mutual information term in~\eqref{eq:equiv-widt-mi} is maximized when $\widetilde{X}_1$ is Gaussian and is independent of $S$. 
Furthermore, by Lemma~\ref{lemma:gaussian-worst},  the  mutual information term on the RHS of~\eqref{eq:lower-bound-isn-ydelta} is  minimized also when  $\widetilde{X}_1$ is Gaussian  and is independent of $S$. 
Therefore, we conclude that 
\begin{IEEEeqnarray}{rCl}
\IEEEeqnarraymulticol{3}{l}{ 
\max_{P_{X_1|S}: \Ex{}{X_1^2} \leq P_1}
\Big\{ I(X_1+S; Y_G) - I(S; Y_\delta)\Big\}
}\notag\\
 &\leq& \max_{b,\rho} \frac{1}{2}\log  \frac{(1+P_2+b +(1+\rho\sqrt{P_1/Q})^2 Q ) (\delta^2 + b)}{(\delta^2 + b + (1+\rho\sqrt{P_1/Q})^2 Q)(1+P_2)} \IEEEeqnarraynumspace
\label{eq:bound-max-gaussian}
\end{IEEEeqnarray}
where the maximization on the RHS is over all pair $(b,\rho)$ satisfying
\begin{IEEEeqnarray}{rCl}
b\geq 0,\quad \text{and}\quad b+\rho^2 P_1 \leq P_1.
\end{IEEEeqnarray}
By examining the Karush-Kuhn-Tucker (KKT) necessary conditions~\cite[Sec. 5.5.3]{boyd04}, it can be shown that the constraint $b+P_1 \rho^2 \leq P_1$ is always binding (namely, the optimal $(b^*,\rho^*)$ pair must satisfy this inequality with equality), and that the optimal $\rho^*$ must be non-positive. 
As a result, the maximization problem on the RHS of~\eqref{eq:bound-max-gaussian} can be simplified to the one dimensional one in~\eqref{eq:def-f-delta}.
The desired bound~\eqref{eq:thm1-ub} follows by substituting~\eqref{eq:def-f-delta} and~\eqref{eq:bound-max-gaussian} into~\eqref{eq:max-single-letter}, then~\eqref{eq:max-single-letter} into~\eqref{eq:ub-r1-max-original},  and by optimizing over~$\delta$.

\subsection{Proof of Proposition~\ref{prop:outer-bound-general}}
\label{sec:proof-prop-outer-nondegraded}
It is straightforward to show the bounds 
\begin{IEEEeqnarray}{rCl}
nR_1 \leq \sum\limits_{i=1}^n I(X_{1,i} ;Y_i|X_{2,i} , S_i ) 
\label{eq:bond-1-time-sharing}
\end{IEEEeqnarray}
and
\begin{IEEEeqnarray}{rCl}
nR_2 \leq \sum\limits_{i=1}^n I(X_{2,i} ;Y_i |X_{1,i} , S_i ) .\label{eq:bond-2-time-sharing}
\end{IEEEeqnarray}
The counterpart of~\eqref{eq:rate-r2-outer-bound-general} can be proved as follows. As in the proof of Theorem~\ref{thm:outer-bound-mac}, we define the rates $R_1$ and $R_2$ as in~\eqref{eq:def-rate-fano-1} and \eqref{eq:def-rate-fano} without loss of generality. 
We have 
\begin{IEEEeqnarray}{rCl}
n(R_1+R_2) &=& I(\msg_1;Y^n) +  I(X_2^n;Y^n) \label{eq:bound-on-R1+R2_1}\\
&=& I(\msg_1, X_2^n;Y_n) - I(X_2^n;M_1|Y^n)\label{eq:bound-on-R1+R2_2}\\
&\leq& h(Y^n) - h(Y^n | \msg_1,X_2^n) \\
&\leq& \sum\limits_{i=1}^n h(Y_i) -h(Y^n | \msg_1,X_2^n) .\label{eq:bound-on-R1+R2_4}
\end{IEEEeqnarray}
Here,~\eqref{eq:bound-on-R1+R2_2} follows because $X_2^n$ and $\msg_1$ are  independent. 
The conditional differential entropy term $h(Y^n | \msg_1,X_2^n) $ can be further lower-bounded as follows:
\begin{IEEEeqnarray}{rCl}
 \IEEEeqnarraymulticol{3}{l}{h(Y^n | \msg_1,X_2^n) }  \\
 &=& h(Y^n, S^n | \msg_1,X_2^n) - h(S^n|Y^n,\msg_1,X_2^n)\\
 &=&h( S^n | \msg_1,X_2^n) + h(Y^n | \msg_1,X_2^n,S^n) \notag\\
 &&-\, h(S^n|Y^n,\msg_1,X_2^n) \\
 &=& h(S^n) + h(Y^n|X_1^n,S^n,X_2^n) - h(S^n|Y^n,\msg_1,X_2^n) \IEEEeqnarraynumspace \label{eq:bound-on-R1+R2_51}\\
&\geq &h(S^n) + h(Y^n|X_1^n,S^n,X_2^n)  - h(S^n|Y^n ,X_2^n) \label{eq:bound-on-R1+R2_5}\\
 &\geq& \sum_{i=1}^n\left(h(S_i) + h(Y_i|X_{1,i},S_i,X_{2,i}) - h(S_i | Y_i,X_{2,i}) \right). \label{eq:bound-on-R1+R2_6} \IEEEeqnarraynumspace
 \end{IEEEeqnarray}
Here, both~\eqref{eq:bound-on-R1+R2_5} and~\eqref{eq:bound-on-R1+R2_6} hold because conditioning does not increase differential entropy. 
Substituting~\eqref{eq:bound-on-R1+R2_6} into~\eqref{eq:bound-on-R1+R2_4}, we conclude that 
\begin{IEEEeqnarray}{rCl}
n(R_1+R_2) &\leq& \sum\limits_{i=1}^{n} \Big(h(Y_i) -  h(Y_i|X_{1,i},S_i,X_{2,i})  \notag\\
&&\qquad -  h(S_i)  +  h(S_i | Y_i,X_{2,i}) \Big)  \\
&=&\sum\limits_{i=1}^{n} \Big(h(Y_i) -  h(Y_i|X_{1,i},S_i,X_{2,i})  \notag\\
&&\qquad -  h(Y_i|X_{2,i})  +  h(Y_i | S_i,X_{2,i}) \Big) \label{eq:bound-sum-singlett} \\
& = & \sum\limits_{i=1}^{n} \Big(I(X_{1,i};Y_i|X_{2,i},S_i) + I(X_{2,i}; Y_i)\Big). \IEEEeqnarraynumspace
\end{IEEEeqnarray}
Here,~\eqref{eq:bound-sum-singlett} follows because $S_i$ and $X_{2,i}$ are independent. 

Introducing the time-sharing random variable $Q$, which is uniformly distributed over the integers $\{1,\ldots,n\}$, we obtain the following outer bound
\begin{IEEEeqnarray}{rCl}
R_1&\leq& I(X_{1};Y|X_2,S,Q) \label{eq:rate-r2-outer-bound1-general-ts}\\
R_2 &\leq& I(X_2;Y|X_1,S,Q)\label{eq:rate-r2-outer-bound2-general-ts}\\
R_1+R_2 &\leq& I(X_1;Y|X_2,S,Q) + I(X_2;Y|Q). \IEEEeqnarraynumspace
\label{eq:rate-r2-outer-bound-general-timesharing}
\end{IEEEeqnarray}
Using the  concavity of mutual information and the fact that $Q$ is independent of $S$, it can be shown  that the above region is equivalent to the one stated in the proposition (without the time sharing random variable $Q$). This concludes the proof.

\subsection{Proof of Theorem~\ref{thm:outer-bound-mac-degraded-1}}
\label{sec:proof-corner-degraded}
 The proof uses techniques similar to the ones used in the proof of Theorem~\ref{thm:outer-bound-mac}. 
The main twist in this case compared with Theorem~\ref{thm:outer-bound-mac} is that $X_2^n$ and $X_1^n$ are not independent. 
To circumvent this, we need to modify the steps in~\eqref{eq:bound-KL-x2-G}--\eqref{eq:ratio-N_S-bound} by conditioning on $M_1$, and by using the fact that $X_1^n$ and $X_2^n$ are conditionally independent given $M_1$.
%
%
%
%
%
%
%
%
%
In particular, the counterpart of $I_\delta$ in~\eqref{eq:def-g-difference-two-mutual-info} is defined as
\begin{IEEEeqnarray}{rCl}
\tilde{I}_{\delta} &\define&  I(X_1^n+S^n; Y_\delta^n|M_1) - I(X_1^n+S^n; Y^n_G|M_1) \IEEEeqnarraynumspace\\
 &=& \frac{n}{2}\Ex{M_1}{\log\frac{\widetilde{N}_S(\delta|M_1)}{\widetilde{N}_S(1+P_2|M_1)} }+ \frac{n}{2}\log \frac{1+P_2}{\delta}
\label{eq:def-I-delta-n-tt}
\end{IEEEeqnarray}
where 
\begin{IEEEeqnarray}{rCl}
\widetilde{N}_{S}(\gamma  | m) \define \exp\lefto\{\frac{2}{n} h(X_1^n +S^n +\sqrt{\gamma} Z^n|M_1=m)\right\}. \IEEEeqnarraynumspace
\end{IEEEeqnarray}
The function $\widetilde{N}_{S}(\gamma  | m) $ inherits all the properties of $N_S(\gamma)$ that are used in Section~\ref{sec:proof-thm-polywu}, such as monotonicity and concavity. 
In the remaining part of the proof,  we omit the mechanical details and only highlight the steps that differ from the ones in Section ~\ref{sec:proof-thm-polywu}.


As in Section~\ref{sec:proof-thm-polywu}, we first upper-bound $\tilde{I}_\delta$.
Let 
\begin{IEEEeqnarray}{rCl}
R_1 &\define& I(M_1;Y^n)\label{eq:def-rate-11-deg}\\
R_2 &\define&  I(X_2^n;Y^n|M_1).\label{eq:def-rate-12-deg}
\end{IEEEeqnarray}
Again, by Fano's inequality, the definitions of the rates in~\eqref{eq:def-rate-11-deg} and~\eqref{eq:def-rate-12-deg} agree with the operational ones. 
With the conditioning on $M_1$, the bounds~\eqref{eq:bound-KL-x2-G} and~\eqref{eq:express-r2-four-entrory} become 
\begin{IEEEeqnarray}{rCl}
D(P_{X_2^n +Z^n |M_1} \| P_{G^n +Z^n} |P_{M_1}) \leq n (C_2 - R_2)\IEEEeqnarraynumspace
\label{eq:bound-KL-x2-G-degraded}
\end{IEEEeqnarray}
and
\begin{IEEEeqnarray}{rCl}
nR_2 
&=&h(Y^n|M_1)- h(Y^n_G|M_1)   \notag\\
&&+\, \Ex{M_1}{\frac{n}{2}\log\frac{\tilde{N}_S(1+P_2|M_1)}{\tilde{N}_S(1|M_1)}}.
\label{eq:express-r2-four-entrory-degraded}
\end{IEEEeqnarray}
Here, $D(P_{X_2^n +Z^n |M_1} \| P_{G^n +Z^n} |P_{M_1}) $ denotes the conditional relative entropy 
\begin{IEEEeqnarray}{rCl}
D(P_{X_2^n +Z^n |M_1} \| P_{G^n +Z^n} |P_{M_1})  \define \Ex{M_1}{ D(P_{X_2^n +Z^n |M_1} \| P_{G^n +Z^n}) }.
\end{IEEEeqnarray}
Using~\cite[Props. 1 and 2]{polyanskiy16-07a} and~\eqref{eq:bound-KL-x2-G-degraded}, we bound the difference $h(Y^n|M_1) -h(Y_G^n | M_1)$ as follows:
\begin{IEEEeqnarray}{rCl}
\IEEEeqnarraymulticol{3}{l}{
h(Y^n|M_1) - h(Y_G^n | M_1) }\notag\\
&\leq& \frac{\log e}{1+P_2}\mathbb{E}_{M_1}\mathopen{}\bigg[ W_2(P_{Y_G^n|M_1},P_{Y^n|M_1}) \bigg(
4\Ex{}{\| X_1^n +S^n\| |M_1}  \notag\\
&& \qquad\qquad\quad +\, \frac{3}{2}\sqrt{\Ex{}{\|Y_G^n\|^2 |M_1} } +\frac{3}{2} \sqrt{\Ex{}{\|Y^n\|^2 |M_1} }\bigg)\bigg] 
\label{eq:cond-M1-optimal-coupling} \\
&\leq&  \frac{\log e}{1+P_2}\mathbb{E}_{M_1}\mathopen{}\bigg[ \sqrt{\frac{2(1+P_2)}{\log e}D(P_{X_2^n +Z^n |M_1} \| P_{G^n +Z^n})} \bigg(
4\sqrt{\Ex{}{\| X_1^n +S^n\|^2 |M_1} } \notag\\
&& \qquad\qquad\quad +\, \frac{3}{2}\sqrt{\Ex{}{\|Y_G^n\|^2 |M_1} } +\frac{3}{2} \sqrt{\Ex{}{\|Y^n\|^2 |M_1} }\bigg)\bigg]  \label{eq:cond-M1-optimal-coupling2} \\
&\leq& \frac{\log e}{1+P_2} \sqrt{\frac{2(1+P_2)}{\log e}D(P_{X_2^n +Z^n |M_1} \| P_{G^n +Z^n} | P_{M_1})}  \notag\\
&&\qquad \cdot \, \left( 4\sqrt{\Ex{}{\| X_1^n +S^n\|^2} } + \frac{3}{2}\sqrt{\Ex{}{\|Y_G^n\|^2 }} + \frac{3}{2} \sqrt{\Ex{}{\|Y^n\|^2}} \right)   \label{eq:cond-M1-optimal-coupling3} \\
&\leq&  c_2 n \sqrt{C_2 -R_2}  \label{eq:cond-M1-optimal-coupling4} 
\end{IEEEeqnarray}
where
\begin{IEEEeqnarray}{rCl}
c_2\define \frac{3\sqrt{1+(\sqrt{P_1}+\sqrt{P_2}+\sqrt{Q})^2} + 4(\sqrt{P_1} + \sqrt{Q}) }{\sqrt{(1+P_2)/(2\log e)}}. \IEEEeqnarraynumspace
\label{eq:def-c1}
\end{IEEEeqnarray}
Here,~\eqref{eq:cond-M1-optimal-coupling} follows from~\cite[Props. 1 and 2]{polyanskiy16-07a};~\eqref{eq:cond-M1-optimal-coupling2} follows because for every message $m$,
\begin{IEEEeqnarray}{rCl}
\Ex{}{\| X_1^n +S^n\| |M_1=m}  \leq \sqrt{\Ex{}{\| X_1^n +S^n\|^2 |M_1=m} } 
\end{IEEEeqnarray}
and
\begin{IEEEeqnarray}{rCl}
W_2(P_{Y_G^n|M_1=m},P_{Y^n|M_1=m}) &\leq& W_2(P_{X_2^n +Z^n |M_1=m} , P_{G^n +Z^n}) \label{eq:w2-yg-yn-1-deg} \\
&\leq&  \sqrt{\frac{2(1+P_2)}{\log e}D(P_{X_2^n +Z^n |M_1=m} \| P_{G^n +Z^n})}\label{eq:w2-yg-yn-2-deg}
\end{IEEEeqnarray}
where~\eqref{eq:w2-yg-yn-1-deg} follows because the $W_2(\cdot,\cdot)$ distance is non-decreasing under convolutions and because $X_1^n+S^n$ and $X_2^n$ are conditionally independent given $M_1$, and the bound~\eqref{eq:w2-yg-yn-2-deg} follows from Talagrand's inequality~\cite{talagrand1996-05a};~\eqref{eq:cond-M1-optimal-coupling3} follows from the Cauchy-Schwarz inequality; and finally~\eqref{eq:cond-M1-optimal-coupling4}  follows from~\eqref{eq:bound-KL-x2-G-degraded},~\eqref{eq:bound-x1n-norm}, and because 
\begin{IEEEeqnarray}{rCl}
\frac{1}{n}\Ex{}{\|Y^n\|^2} &\leq& 1+(\sqrt{P_1}+\sqrt{P_2}+\sqrt{Q})^2\\
\frac{1}{n}\Ex{}{\|Y^n_G\|^2}  &\leq& 1+(\sqrt{P_1}+\sqrt{P_2}+\sqrt{Q})^2.
\end{IEEEeqnarray}
Substituting~\eqref{eq:cond-M1-optimal-coupling4}   into~\eqref{eq:express-r2-four-entrory-degraded}, we conclude that
\begin{equation} 
\Ex{M_1}{\log\frac{\tilde{N}_S(1+P_2|M_1)}{\tilde{N}_S(1|M_1)}} \leq 2 c_2\sqrt{ C_2-R_2 }  +  2(C_2 - R_2) -\log(1+P_2).
\label{eq:bound-on-ratio-ns-initial}
\end{equation}
Letting $\alpha \define P_2/(1+P_2 -\delta)$ as in  Section~\ref{sec:proof-thm-polywu}, we obtain 
\begin{IEEEeqnarray}{rCl}
\IEEEeqnarraymulticol{3}{l}{
\Ex{M_1}{\log\frac{\widetilde{N}_S(\delta|M_1)}{\widetilde{N}_S(1+P_2|M_1)} } }\notag\\
&\leq & \Ex{M_1}{ \log \lefto( \frac{\widetilde{N}_S(1|M_1) }{\widetilde{N}_S(1+P_2|M_1) } -1+ \alpha \right)} -\log \alpha \label{eq:bound-cond-entropy-pwer1}\\
&\leq&  \log \left( \frac{\exp\lefto(2 c_2\sqrt{C_2-R_2} + 2(C_2-R_2) \right) }{1+P_2} - 1 +\alpha \right) -\log \alpha.
\label{eq:ratio-N_S-bound-degraded}
\end{IEEEeqnarray}
Here,~\eqref{eq:bound-cond-entropy-pwer1} follows from the concavity of $\gamma\mapsto \widetilde{N}_{S}(\gamma  | M_1) $, 
and~\eqref{eq:ratio-N_S-bound-degraded} follows from Jensen's inequality and because  the function $x\mapsto \log(\exp(x) -(1-\alpha))$ is concave. 
Finally, substituting~\eqref{eq:ratio-N_S-bound-degraded}  into~\eqref{eq:def-I-delta-n-tt}, we conclude that 
\begin{IEEEeqnarray}{rCl}
\tilde{I}_\delta &\leq& \frac{n}{2} \log\lefto( 1 + \frac{1+P_2-\delta}{ P_2\delta} \tilde{g}(R_2) \right) 
\label{eq:upper-bound-I-delta-final-degraded}
\end{IEEEeqnarray}
where $\tilde{g}(R_2)$ is defined in~\eqref{eq:def-g-r2-degraded}.

We next relate $\tilde{I}_\delta$ to $R_1$. This part is quite different from the steps in Section~\ref{sec:step2-lower-bound-idelta}, since for the dirty MAC with degraded message sets,  the information about the message $M_1$ is contained in both  $X_1^n$ and $X_2^n$. 
Consider the following chain:
\begin{IEEEeqnarray}{rCl}
\tilde{I}_{\delta} &=& I(X_1^n,S^n;Y^n_\delta |M_1) - I(X_1^n+S^n,M_1;Y_G^n) + I(M_1;Y^n_G) \\
&=&I(S^n;Y^n_{\delta},M_1)  + I(X_1^n;Y^n_{\delta}|S^n,M_1) - I(X_1^n+S^n,M_1;Y_G^n) + I(M_1;Y^n_G) \\
&\geq& I(S^n;Y_\delta^n)- I(X_1^n+S^n;Y_G^n) + I(M_1;Y^n_G) \\
&=& I(S^n;Y_\delta^n)- I(X_1^n+S^n;Y_G^n)  + I(M_1;Y_G^n) - I(M_1;Y^n) + nR_1.  \label{eq:lower-bound-tI-delta-1}
\end{IEEEeqnarray}
Here, the penultimate  step follows because $M_1\to X_1^n+S^n \to Y_G^n$ forms a Markov chain. 
The first two terms on the RHS of~\eqref{eq:lower-bound-tI-delta-1} can be single-letterized and bounded in the same way as in Section~\ref{sec:single-letterization-proof-thm1} and Section~\ref{sec:gaussian-optimal-proof-th1}, i.e.,
\begin{IEEEeqnarray}{rCl}
I(S^n;Y_\delta^n)- I(X_1^n+S^n;Y_G^n) \geq -n f(\delta)
\label{eq:diff-sn-ydelta-n-g-n-degraded}
\end{IEEEeqnarray}
where $f(\cdot)$ was defined in~\eqref{eq:def-f-delta}.

To conclude the proof, it remains to lower-bound $I(M_1;Y_G^n) - I(M_1;Y^n)$. To this end, we rewrite it as 
\begin{IEEEeqnarray}{rCl}
I(M_1;Y_G^n) - I(M_1;Y^n) = h(Y_G^n) - h(Y^n) + h(Y^n|M_1) -h(Y^n_G|M_1).
\label{eq:bound-diff-m1-ygn}
\end{IEEEeqnarray}
The differences $ h(Y_G^n) - h(Y^n) $ and $h(Y^n|M_1) -h(Y^n_G|M_1)$ can be bounded via steps similar to those  in~\eqref{eq:cond-M1-optimal-coupling}--\eqref{eq:cond-M1-optimal-coupling4}. More specifically, we have 
\begin{IEEEeqnarray}{rCl}
h(Y^n_G|M_1) - h(Y^n|M_1) \leq c_3 n \sqrt{C_2 -R_2}
\label{eq:bound-diff-hynm1-2}
\end{IEEEeqnarray}
and
\begin{IEEEeqnarray}{rCl}
h(Y^n) - h(Y^n_G)  \leq c_2 n\sqrt{C_2-R_2}
\label{eq:diff-hyn-yng}
\end{IEEEeqnarray}
where $c_3$ was defined in~\eqref{eq:def-c3}. 
Here, to prove~\eqref{eq:diff-hyn-yng}, we have used 
\begin{IEEEeqnarray}{rCl}
D(P_{Y^n}\| P_{Y_G^n}) &\leq&  D(P_{Y^n|M_1} \| P_{Y_G^n |M_1} | P_{M_1})  \label{eq:diff-hyn-yng-1}\\
&\leq& D(P_{X_2^n +Z^n |M_1} \| P_{G^n +Z^n} |P_{M_1})\label{eq:diff-hyn-yng-2} \\
&\leq& n (C_2 - R_2)\label{eq:diff-hyn-yng-3}
\end{IEEEeqnarray}  
where~\eqref{eq:diff-hyn-yng-1} follows from the data processing inequality,~\eqref{eq:diff-hyn-yng-2} follows from the data processing inequality and because 
$X_1^n+S^n$ and $X_2^n$ are conditionally independent given $M_1$, and~\eqref{eq:diff-hyn-yng-3} follows from~\eqref{eq:bound-KL-x2-G-degraded}. 
Substituting~\eqref{eq:bound-diff-hynm1-2} and~\eqref{eq:diff-hyn-yng} into~\eqref{eq:bound-diff-m1-ygn}, then~\eqref{eq:bound-diff-m1-ygn} and~\eqref{eq:diff-sn-ydelta-n-g-n-degraded} into~\eqref{eq:lower-bound-tI-delta-1}, and combining~\eqref{eq:lower-bound-tI-delta-1} with~\eqref{eq:upper-bound-I-delta-final-degraded}, we conclude the proof of~\eqref{eq:thm1-ub-degraded}.

\subsection{Proof of Proposition~\ref{prop:outer-bound-degraded}}
\label{prop-outer-bound-degraded-mac-sum}
The key idea of the proof is to identify the auxiliary random variables $U\define (M_1, Q)$, where~$Q$ denotes the time-sharing random variable that is uniformly distributed over the integers $\{1,\ldots,n\}$. 
We have 
\begin{IEEEeqnarray}{rCl}
nR_2 &=& I(X_2^n;Y^n|M_1) \\
&\leq& I(X_2^n;Y^n, X_1^n, S^n|M_1)\\
&= &I(X_2^n;Y^n|X_1^n, S^n , M _1)\\
&=& h(Y^n|X_1^n, S^n , M _1) -h(Y^n|X_1^n,X_2^n ,S^n,M_1)\\
&\leq& \sum\limits_{i=1}^{n} h(Y_i|X_{1,i},S_{i},M_1) - h(Y_i| X_{1,i},X_{2,i},S_i,M_1)\IEEEeqnarraynumspace\\
&=&\sum\limits_{i=1}^{n} I(X_{2,i};Y_{i} | X_{1,i} ,S_{i}, M_1)\\
&=&I(X_{2};Y | X_1 ,S, U).
\end{IEEEeqnarray}
This yields the upper bound in~\eqref{eq:out-prop-degraded-r2}.

To prove~\eqref{eq:out-prop-degraded-r22}, we observe that  
\begin{IEEEeqnarray}{rCl}
R_2 &=& I(X_2^n;Y^n|M_1) \\
&=& h(Y^n|M_1) - h(Y^n|M_1,X_2^n) \\
&\leq& \sum\limits_{i=1}^{n} h(Y_i |M_1) - h(Y^n|M_1,X_2^n).
\end{IEEEeqnarray}
Proceeding as in~\eqref{eq:bound-on-R1+R2_4}--\eqref{eq:rate-r2-outer-bound-general-timesharing} while keeping the conditioning on $M_1$, we conclude that  
\begin{IEEEeqnarray}{rCl}
R_2& \leq & \sum\limits_{i=1}^{n} \Big(I(X_{1,i};Y_i|X_{2,i},S_i,M_1) + I(X_{2,i}; Y_i |M_1)\Big)  \IEEEeqnarraynumspace \\
&=& I(X_1;Y|X_2,S,M_1,Q) +I(X_2;Y|Q,M_1) \\
&\leq&  I(X_1;Y|X_2,S,U) + I(X_2;Y|U).
\end{IEEEeqnarray}

Finally, we prove~\eqref{eq:out-sum-degraded}.  We proceed again as in~\eqref{eq:bound-on-R1+R2_1}--\eqref{eq:bound-on-R1+R2_6} and  keep the conditioning on $M_1$ whenever appropriate. This yields 
\begin{IEEEeqnarray}{rCl}
\IEEEeqnarraymulticol{3}{l}{n(R_1+R_2)} \notag\\
&\leq& \sum\limits_{i=1}^{n}\Big( h(Y_i) - h(Y_i|X_{1,i},S_i,X_{2,i})\Big)\notag\\
&&-\,  h(S^n|M_1)  + h(S^n|Y^n,\msg_1,X_2^n) \\
&\leq & \sum\limits_{i=1}^{n} \Big(h(Y_i) -  h(Y_i|X_{1,i},S_i,X_{2,i},M_1)  \notag\\
&&\qquad - \, h(S_i|M_1,X_{2,i})  +  h(S_i | M_1 ,Y_i,X_{2,i}) \Big)  \label{eq:bound-sum-degraded-3}\\
& = & \sum\limits_{i=1}^{n} \Big(I(X_{1,i};Y_i|X_{2,i},S_i,M_1) + I(X_{2,i},M_1; Y_i)\Big)  \IEEEeqnarraynumspace \\
&=& I(X_1;Y|X_2,S,M_1,Q) +I(X_2,M_1;Y|Q) \\
&\leq&  I(X_1;Y|X_2,S,U) + I(X_2,U;Y).
\end{IEEEeqnarray}
Here,~\eqref{eq:bound-sum-degraded-3} follows because $S_i$ is independent of $M_1$ and $X_{2,i}$, and because conditioning does not increase entropy.
The proof is concluded by observing that the auxiliary random variable $U$ and the random variables $X_1$, $X_2$, $S$ satisfy the conditions listed in the theorem. 

\section{Conclusion}

In this paper, we have studied a two-user state-dependent Gaussian MAC with state noncausally known at one encoder and with and without degraded message sets. 
We have derived several new outer bounds on the capacity region, which provide substantial improvements over the best previously known outer bounds. 
For the dirty MAC without degraded message sets, our outer bounds yield the following:
\begin{itemize}
\item The characterization of the sum rate capacity;
\item The establishment of the two corner points of the capacity region;
\item The characterization of the full capacity region in the special case  in which the sum rate capacity is equal to the capacity $C_{\mathrm{helper}}$ of the helper problem;
\item A new upper bound on $C_{\mathrm{helper}}$, and a necessary and sufficient condition to achieve $C_{\mathrm{helper}} = \frac{1}{2}\log(1+P_2)$.
\end{itemize}
We have shown that a single-letter solution is adequate to achieve both the corner points and the sum rate capacity.
In addition, we have  generalized our outer bounds to the case of  additive non-Gaussian states. 
%
%


 There are several possible generalizations of the results in this paper. 
 \begin{itemize}
 \item The outer bounds derived in this paper can be readily generalized  to the discrete  and to the multiple-input multiple-output (MIMO) setting. This is unlike the \emph{doublely dirty Gaussian MAC setting},  in which additional difficulties arise when extending from the single-input single-output to the MIMO setting~\cite{Khina2017-05}.
\item In this paper, we assume that the state is not known at the non-cognitive user.  
It would be interesting to investigate whether revealing the state information   strictly
causally to the non-cognitive user  can increase the capacity region.
As shown in~\cite{lapidoth2013-01a}, strictly causal state information enables cooperations between the two encoders (e.g., by letting the encoders convey the past state information jointly  to the decoder). 

 \item In the proofs of Theorem~\ref{thm:outer-bound-mac} and Theorem~\ref{thm:outer-bound-mac-degraded-1}, we have essentially transformed the dirty MAC into a state-dependent $Z$-interference channel with input-output relationship 
 \begin{IEEEeqnarray}{rCl}
Y_1 &=& X_1+S+\sqrt{\delta} Z_1\\
 Y_2 &=& X_1+ X_2 +S+Z_2 
\end{IEEEeqnarray}
 where the Gaussian noises $Z_1,Z_2 \sim \mathcal{N}(0,1)$ are independent. 
 This suggests that our techniques may yield tighter outer bounds on the capacity region of the state-dependent Gaussian $Z$-interference channel than the ones derived in~\cite{duan2015-12a}.  

 \item Another related setting is the state-dependent relay channel with state available noncausally at the relay considered in~\cite{zaidi2010-05a}.
  It would be interesting to see whether our  techniques can lead to any improvement over the bounds there. 
 \end{itemize}

\begin{appendix}[Gaussian Inputs Maximize \eqref{eq:out-prop-degraded-r2}--\eqref{eq:out-sum-degraded}]
\label{sec:gaussian-optimality}
We shall prove that the outer region provided in Proposition~\ref{prop:outer-bound-degraded} is maxmized when $U$, $S$, $X_1$, and $X_2$ are jointly Gaussian distributed. 
Differently from~\cite[Th.~4]{zaidi2009-06a}, the presence of the auxiliary random variable $U$ complicates the proof substantially. 
Consider an arbitrary distribution $P_{USX_1X_2}$ that satisfies the conditions stated in the proposition. 
Without loss of generality, we assume that $P_{USX_1X_2}$ satisfies the following conditions, in addition to the ones stated in Proposition~\ref{prop:outer-bound-degraded}:
\begin{itemize}
\item $U$ has zero mean and unit variance;
\item $\Ex{}{X_1^2} =P_1$ and $\Ex{}{X_2}=P_2$.
\end{itemize} 
The first assumption comes without loss of generality since  $U$ does not appear in the channel input-output relation $Y=X_1+X_2+S+Z$, and the second assumption comes without loss of generality because we do not assume $X_1$ and $X_2$ to have  zero mean.
We next introduce the following notation:
\begin{IEEEeqnarray}{rCl}
\mu_{k}(u) &\define& \Ex{}{X_k|U=u} \\
\sigma_{k}(u) &\define& \sqrt{\mathrm{Var}[X_k|U=u]} \label{eq:def-sigma-k-deg}\\
\rho_k &\define & \sqrt{\Ex{}{\mu_k^2(U)}/P_k}\\
\mu_{s}(u) &\define& \Ex{}{X_1S|U=u}/\sqrt{Q} \label{eq:def-mus-deg} \\
\rho_s &\define& \Ex{}{\mu_s(U)}/\sqrt{P_1}  
\end{IEEEeqnarray}
where $k\in\{1,2\}$.
It follows that 
\begin{IEEEeqnarray}{rCl}
R_1 &\leq& I(X_2;Y|X_1,S,U) \\
&\leq& \frac{1}{2}\Ex{}{\log(1+ \sigma_2(U)^2  } \\
&\leq& \frac{1}{2}\log\lefto(1+ \Ex{}{\sigma_2(U)^2 } \right)\label{eq:bound-r1-degraded-step3}\\
&=& \frac{1}{2}\log(1+ P_2(1-\rho_2^2)).\label{eq:bound-r1-degraded-step4}
\end{IEEEeqnarray}
Here,~\eqref{eq:bound-r1-degraded-step3} follows from Jensen's inequality, and~\eqref{eq:bound-r1-degraded-step4} follows because 
\begin{IEEEeqnarray}{rCl}
\Ex{}{\sigma^2_2(U) }= \Ex{}{ \Ex{}{X_2^2|U} - \mu_2(U)^2} = P_2 - \rho_2^2P_2. \IEEEeqnarraynumspace
\end{IEEEeqnarray}
This proves~\eqref{eq:out-r2-degraded-gau}.

To prove~\eqref{eq:out-r2-degraded-gau2}, we proceed as follows:
\begin{IEEEeqnarray}{rCl}
R_2&\leq& I(X_1;Y|X_2,S,U) + I(X_2;Y|U) \\
&=& I(X_1,X_2,S;Y|U) - I(S;Y|U,X_2) .  \label{eq:out-prop-degraded-r22-proof} \IEEEeqnarraynumspace
\end{IEEEeqnarray}
To upper-bound $I(X_1,X_2,S;Y|U)$, we observe that  
\begin{IEEEeqnarray}{rCl}
\IEEEeqnarraymulticol{3}{l}{\mathrm{Var}[X_1+X_2+S| U=u]}\notag\\ 
&=& \sigma_1^2(u) +\sigma_2^2(u) + Q+ 2\sqrt{Q}\mu_s(u) 
\end{IEEEeqnarray}
where we have used~\eqref{eq:def-sigma-k-deg} and~\eqref{eq:def-mus-deg}, and that   $X_1$ and $X_2$ are conditionally independent given~$U$.
It thus follows that 
\begin{IEEEeqnarray}{rCl}
\IEEEeqnarraymulticol{3}{l}{
 I(X_1+X_2+S;Y|U) }\notag\\
&\leq &\frac{1}{2}\Ex{}{\log(1+\sigma_1^2(U) +\sigma_2^2(U) +Q+ 2\sqrt{Q}\mu_s(U) )}  \IEEEeqnarraynumspace
\\ 
&\leq &\frac{1}{2} \log\lefto(1+ \Ex{}{\sigma_1^2(U) +\sigma_2^2(U)+Q+ 2\sqrt{Q}\mu_s(U) } \right) 
\IEEEeqnarraynumspace\\
&=& \frac{1}{2} \log\lefto( 1+ P_1(1-\rho_1^2) +P_2(1-\rho_2^2) +Q+  2 \rho_s \sqrt{QP_1}  \right) . 
\label{eq:bound-hy-hz-degraded-cond} 
\end{IEEEeqnarray}
Here, in~\eqref{eq:bound-hy-hz-degraded-cond} we have used the following identity:
\begin{IEEEeqnarray}{rCl}
\Ex{}{\sigma^2_k(U)} &=& \Ex{}{\mathrm{Var}[X_k|U]} \\
&=& \mathrm{Var}[X_k] - \mathrm{Var}[\mu_k(U)] \label{eq:law-total-variance}\\
&=& \mathrm{Var}[X_k] - \Ex{}{\mu_k(U)^2} + \Ex{}{X_k}^2 \\
&=&P_k - P_k\sigma_k^2,\quad k\in\{1,2\}\label{eq:law-total-variance-2}
\end{IEEEeqnarray}
where~\eqref{eq:law-total-variance} follows from the law of total variance.

We next bound the second term on the RHS of~\eqref{eq:out-prop-degraded-r22-proof}. Let  
\begin{IEEEeqnarray}{rCl}
\widetilde{X}_1 \define X_1 - \mu_1(U) - \frac{\mu_s(U)S}{\sqrt{Q}} .
\label{eq:def-widt-x1-deg}
\end{IEEEeqnarray}
It follows that 
\begin{IEEEeqnarray}{rCl}
\Ex{}{\widetilde{X}_1S|U=u} = \Ex{}{X_1S|U=u} -  \mu_s(u) \sqrt{Q} =0. \IEEEeqnarraynumspace
\label{eq:zero-mean-givenu-x1s}
\end{IEEEeqnarray}
Since $S$ is Gaussian distributed, by Lemma~\ref{lemma:gaussian-worst},
\begin{IEEEeqnarray}{rCl}
\IEEEeqnarraymulticol{3}{l}{
 I(S;Y| X_2, U) }\\
 &=& \Ex{U}{I(S; (1 + \mu_s(U)/\sqrt{Q}) S + \widetilde{X}_1 + Z |U) } \IEEEeqnarraynumspace\\
  &\geq&  \frac{1}{2}\Ex{}{\log\left(1 + \frac{(\sqrt{Q} +\mu_s(U) )^2 }{1 + \sigma_1^2(U) -   \mu_s(U)^2 }  \right)}. 
\end{IEEEeqnarray}
By~\eqref{eq:def-sigma-k-deg},~\eqref{eq:def-widt-x1-deg}, and~\eqref{eq:zero-mean-givenu-x1s},
\begin{IEEEeqnarray}{rCl}
\sigma_1^2(u) &=&\Ex{}{X_1^2 |U=u} - \mu_1^2(u)\\
 &=& \Ex{}{\widetilde{X}_1^2|U=u} +   \mu_s(u)^2 \geq \mu_s(u)^2.\IEEEeqnarraynumspace
 \label{eq:inequality-bound-sigma1-rhos}
\end{IEEEeqnarray}
Now, observe that the function 
\begin{IEEEeqnarray}{rCl}
\xi(a,b) \define \frac{1}{2} \log\lefto( 1+ \frac{(\sqrt{Q} -a)^2}{1+b-a^2}\right) 
\end{IEEEeqnarray}
is jointly convex in $(a,b)$ as long as  $a^2\leq b$. Indeed, let $\matH$ be the Hessian matrix of $\xi(a,b)$. It follows that 
\begin{IEEEeqnarray}{rCl}
H_{11} &=& \frac{\partial^2 \xi}{\partial a^2} \\
&=& \frac{(\sqrt{Q} -a)^2 ((\sqrt{Q} -a)^2 + 2 + 2b- 2a^2)}{( 1+ b-a^2)^2( \sqrt{Q} -a)^2 + 1+ b-a^2 )^2 } \IEEEeqnarraynumspace\\
&\geq& 0
\end{IEEEeqnarray}
and that 
\begin{IEEEeqnarray}{rCl}
\mathrm{Det}[\matH] &=& \frac{ (\sqrt{Q} -a)^4}{( 1+ b-a^2)^3( \sqrt{Q} -a)^2 + 1+ b-a^2 )^2 } \IEEEeqnarraynumspace \\
& \geq & 0. 
\end{IEEEeqnarray}
Therefore, $\matH$ is positive semi-definite for all $(a,b)$ satisfying $a^2 \leq b$, which implies that  the function $\xi(a,b) $ is convex. 
Therefore, by Jensen's inequality, 
\begin{IEEEeqnarray}{rCl}
\IEEEeqnarraymulticol{3}{l}{
 I(S;X_1+S+Z| U) }\\
  &\geq&  \frac{1}{2}\log\left(1 + \frac{(\sqrt{Q} +  \Ex{}{\mu_s(U)})^2 }{1 + \Ex{}{\sigma_1^2(U)} - \Ex{}{\mu_s(U)}^2 }  \right) \\
  &=&  \frac{1}{2}\log\left(1 + \frac{(\sqrt{Q} + \rho_s \sqrt{P_1})^2 }{1 + P_1 -   \rho_1^2 P_1 -\rho_s^2P_1 }  \right) .\label{eq:bound-i-cond=mu}
\end{IEEEeqnarray}
Here, in~\eqref{eq:bound-i-cond=mu} we have used~\eqref{eq:law-total-variance-2}.
Substituting~\eqref{eq:bound-hy-hz-degraded-cond} and~\eqref{eq:bound-i-cond=mu} into~\eqref{eq:out-prop-degraded-r22-proof} and rearranging the terms, we obtain~\eqref{eq:out-r2-degraded-gau2}.

The proof of~\eqref{eq:out-sum-degraded-gau} follows  steps analogous to those in  the proof of~\eqref{eq:out-r2-degraded-gau2}.
More specifically, we obtain from~\eqref{eq:out-sum-degraded} that 
\ifthenelse{\boolean{singcol}}{
\begin{IEEEeqnarray}{rCl}
R_1 +  R_2   &\leq& h(Y|X_2,S,U)- h(Y|X_1,X_2,S,U) \notag\\
&&+\, h(Y) - h(Y|X_2, U) \\
&=& I(X_1+X_2+S;Y)  - I(S;X_1+S+Z| U)  . \IEEEeqnarraynumspace
\label{eq:general-bound-sum-degraded}
\end{IEEEeqnarray} 
}{
\begin{IEEEeqnarray}{rCl}
\IEEEeqnarraymulticol{3}{l}{
R_1 +  R_2}\notag\\
\quad  &\leq& h(Y|X_2,S,U)- h(Y|X_1,X_2,S,U) \notag\\
&&+\, h(Y) - h(Y|X_2, U) \\
&=& I(X_1+X_2+S;Y)  - I(S;X_1+S+Z| U)  . \IEEEeqnarraynumspace
\label{eq:general-bound-sum-degraded}
\end{IEEEeqnarray}} 
The term $I(S;X_1+S+Z| U) $ on the RHS of~\eqref{eq:general-bound-sum-degraded} has been lower-bounded in~\eqref{eq:bound-i-cond=mu}. 
To upper-bound $I(X_1+X_2+S;Y)  $, we bound $\Ex{}{(X_1+X_2+S)^2}$ as 
\begin{IEEEeqnarray}{rCl}
\IEEEeqnarraymulticol{3}{l}{\Ex{}{(X_1+X_2+S)^2}}\notag\\ 
&=&P_1 + P_2 + Q+ 2\Ex{}{X_1S} + 2\Ex{}{X_1X_2} \\
&=&  P_1 + P_2 + Q + 2\rho_s\sqrt{P_1Q} + 2\Ex{}{\Ex{}{X_1|U }\Ex{}{X_2|U} } \IEEEeqnarraynumspace\label{eq:bound-power-y2-2}\\
&\leq& P_1 +P_2+Q + 2\rho_s\sqrt{P_1Q}  + 2\rho_1\rho_2\sqrt{P_1P_2} . \IEEEeqnarraynumspace \label{eq:bound-power-y2-3}
\end{IEEEeqnarray}
Here,~\eqref{eq:bound-power-y2-2} follows because $X_1$ and $X_2$ are conditionally independent given $U$, 
 and~\eqref{eq:bound-power-y2-3} follows because
\begin{IEEEeqnarray}{rCl}
\Ex{}{\Ex{}{X_1|U }\Ex{}{X_2|U}}   &=& \Ex{}{ \mu_1(U) \mu_2(U) }  \\
&\leq& \sqrt{\Ex{}{ \mu_1(U)^2}  \Ex{}{\mu_2(U)^2}} \IEEEeqnarraynumspace\\
&=&\rho_1\rho_2\sqrt{P_1P_2}.
\end{IEEEeqnarray}
It thus follows that 
\begin{IEEEeqnarray}{rCl}
 I(X_1+X_2+S;Y)  &\leq &\frac{1}{2}\log\mathopen{}\Big(1+P_1 +P_2+Q \notag\\
 &&+ \,2\rho_s\sqrt{P_1Q}  + 2\rho_1\rho_2\sqrt{P_1P_2} \Big).\IEEEeqnarraynumspace
 \label{eq:bound-hy-hz-degraded}
\end{IEEEeqnarray}
Substituting~\eqref{eq:bound-hy-hz-degraded} and~\eqref{eq:bound-i-cond=mu} into~\eqref{eq:general-bound-sum-degraded}, we obtain~\eqref{eq:out-sum-degraded-gau}.

Finally, observe from~\eqref{eq:law-total-variance-2} and~\eqref{eq:inequality-bound-sigma1-rhos}  that 
\begin{equation}
P_1 - P_1\sigma_1^2 = \Ex{}{\sigma^2_1(U)}  \geq \Ex{}{\mu_s(U)^2}\geq \Ex{}{\mu_s(U)}^2 \geq P_1\rho_s^2
\end{equation}
which implies the condition~\eqref{eq:cond-rho1-rhos}. This concludes the proof.

\end{appendix}


%
%

\end{document}